\documentclass[10pt]{article}

\pdfoutput=1

\usepackage{a4wide}
\usepackage{graphicx}

\usepackage{amssymb,amsfonts,amsmath}
\usepackage{marvosym}
\usepackage{wasysym}

\usepackage{hyperref}

\newcommand{\E}{\mbox{\scriptsize \Earth}}

\let\oldepsilon\epsilon
\let\epsilon\varepsilon
\let\varepsilon\oldepsilon

\begin{document}

\begin{center}   
\textbf{\LARGE Giant ripples on comet 67P/Churyumov-Gerasimenko sculpted by sunset thermal wind}

\vspace*{0.2cm}

P. \textsc{Jia}, B. \textsc{Andreotti} and P. \textsc{Claudin}
\end{center}

{\small
\noindent
Laboratoire de Physique et M\'ecanique des Milieux H\'et\'erog\`enes (PMMH),\\
UMR CNRS 7636, ESPCI Paris - PSL Research University, 10 rue Vauquelin, 75005 Paris, France;\\
Univ. P. M. Curie - Sorbonne Universit\'es; Univ. D. Diderot - Sorbonne Paris Cit\'e.
}

\begin{abstract}
Explaining the unexpected presence of dune-like patterns at the surface of the comet 67P/Churyumov-Gerasimenko requires conceptual and quantitative advances in the understanding of surface and outgassing processes. We show here that vapor flow emitted by the comet around its perihelion spreads laterally in a surface layer, due to the strong pressure difference between zones illuminated by sunlight and those in shadow. For such thermal winds to be dense enough to transport grains -- ten times greater than previous estimates -- outgassing must take place through a surface porous granular layer, and that layer must be composed of grains whose roughness lowers cohesion consistently with contact mechanics. The linear stability analysis of the problem, entirely tested against laboratory experiments, quantitatively predicts the emergence of bedforms in the observed wavelength range, and their propagation at the scale of a comet revolution. Although generated by a rarefied atmosphere, they are paradoxically analogous to ripples emerging on granular beds submitted to viscous shear flows. This quantitative agreement shows that our understanding of the coupling between hydrodynamics and sediment transport is able to account for bedform emergence in extreme conditions and provides a reliable tool to predict the erosion and accretion processes controlling the evolution of small solar system bodies.
\end{abstract}

\section{Introduction}
\label{intro}

The OSIRIS imaging instrument on board the ESA's Rosetta spacecraft has revealed unexpected bedforms  (Fig 1 and S1) on the neck of the comet 67P/Churyumov-Gerasimenko (the Hapi region) \cite{ThomasetalSci2015,ElMaarryetal2015,ThomasetalAA2015} and on both lobes (Ma'at and Ash regions). Several features suggest that these rhythmic patterns belong to the family of ripples and dunes \cite{CAC2013}. The bedforms present a characteristic asymmetric profile, with a small steep lee side resembling an avalanche slip face (Figs.~1A, S1B) and a longer gentle slope on the stoss side, which appears darker in Fig.~1B. Analysis of the available photographs show that their typical crest-to-crest distance is on the order of $10$~m (Tab.~S1), and that the surface  is composed of centimeter scale grains \cite{Mottolaetal2015} (Fig.~2). However, the existence of sedimentary bedforms on a comet comes as a surprise -- it requires sediment transport along the surface, i.e. erosion and deposition of particles. When heated by the sun, the ice at the surface of comets sublimates into gas. As gravity is extremely small, $g \simeq 2 \, 10^{-4}$~m/s$^2$, due to the kilometer scale of the comet \cite{Sierksetal2015,Paetzoldetal2016}, the escape velocity is much smaller than the typical thermal velocity. Outgassing therefore feeds an extremely rarefied atmosphere, called the coma, around the nucleus (Fig. 4B). This gas envelope expands radially. By contrast, ripples and dunes observed in deserts, on the bed of rivers and on Mars and Titan \cite{CAC2013,CA2006,Bourkeetal2010,FEBL2013,Lucasetal2014,Letal2016} are formed by fluid flows parallel to the surface, dense enough to sustain sediment transport. The presence of these apparent dunes therefore challenges the common views of surface processes on comets and raises several questions. What could be the origin of the vapor flow exceeding the sediment transport velocity threshold \cite{OCA2011,KPMBK2012}? How could the particles of the bed remain confined to the surface of the comet rather than being ejected into the coma? Our goal here is to understand the emergence of the bedforms on 67P and to constrain the modeling of dynamical processes in the superficial layer of the comet nucleus.

\begin{figure}[p]
\centerline{\includegraphics{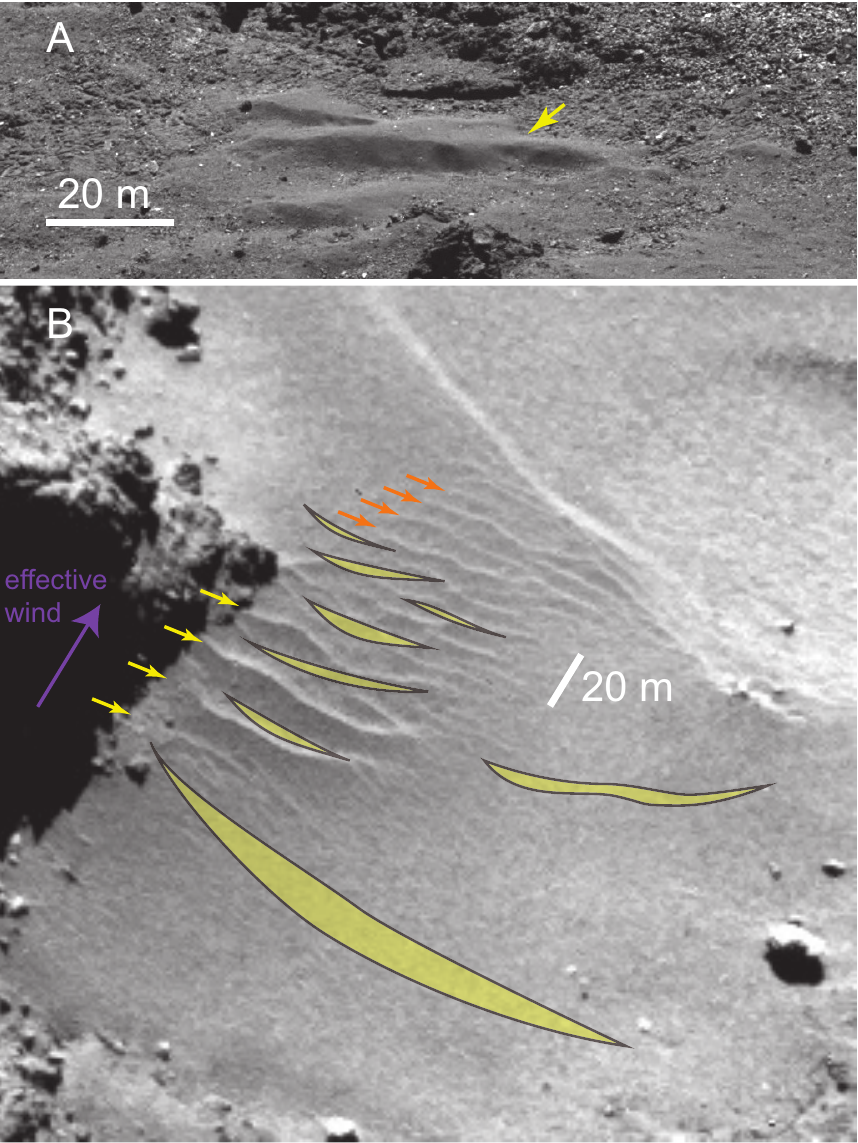}}
\caption{Ripples. \textbf{A} Photograph of ripples in the Maftet region. The central bedform (yellow arrow) has a length $\lambda \simeq 20$~m (see also Fig.~S1b) and a height around $2$~m, i.e. with a typical aspect ratio $0.1$. \textbf{B} View of the comet's bedforms in the neck (Hapi) region by OSIRIS narrow-angle camera dated 18 September 2014, i.e. before perihelion. Superimposed yellow marks (Methods): position of the ripples from a photo dated 17 January 2016 (Fig. S1), i.e. after perihelion providing evidence for their activity. The mean crest-to-crest distance $\lambda$ ranges from  $\simeq 7$~m (emergent ripples upwind of the largest slip face: orange arrows) to $\simeq 18$~m for the larger bedforms (yellow arrows). All Photo credits: ESA/Rosetta/MPS, see Tab.~S1 for references.}
\label{Fig1}
\end{figure}
%

\section{Outgassing and comet's atmosphere}
Outgassing takes place in the illuminated part of the comet \cite{DeSanctisetal2015,Filacchioneetal2016}. As ice sublimation requires an input of energy -- the latent heat -- the vapour flux is controlled by the thermal balance at the surface of the comet (Methods). The power per unit area received from the sun depends, at the seasonal scale, on the heliocentric distance and is modulated by the day-night alternation. The comet radiates some energy back to space with a power related to the surface temperature $T_s$ by Stefan's law. Finally, thermal inertia leads to a storage/release of internal energy over a penetration depth which is meter scale for seasonal variations and centimeter scale for daily variations.

The vapor production rate from outgassing, defined as the product of the vapor density $\rho_0$ by the outward vapor velocity $u_0$, has been measured for 67P at different heliocentric distances \cite{Gulkisetal2015,Hanneretal1985,OSM1992,S2006,OKKUSOSOS2012,BCQS2014} (Fig. 3B). Common models assume that ice sublimation takes place at the surface and produces a radial flow at the thermal velocity \cite{SR1998}. This would result in a density $\rho_0$ an order of magnitude smaller than that necessary to induce a fluid drag force large enough to overcome the threshold for grain motion (as discussed below). We suggest that most of the vapor is emitted from sub-surface ice and must travel through the porous surface granular layer (Fig. S3). Sublimation makes the ice trapped in the pores recede, releasing unglued grains in surface that can be eroded. This process should lead to an ice level remaining at a constant distance from the surface, comparable to the grain size $d$. Using kinetic theory of gasses, we predict that for such vapor flow the outgassing velocity is ten times smaller than that of the spectacular vapor jets streaming from active pits \cite{Sierksetal2015,Vincentetal2015} (Methods). Accordingly, the vapor atmosphere is ten times denser than previous estimates.

\begin{figure}[p]
\centerline{\includegraphics{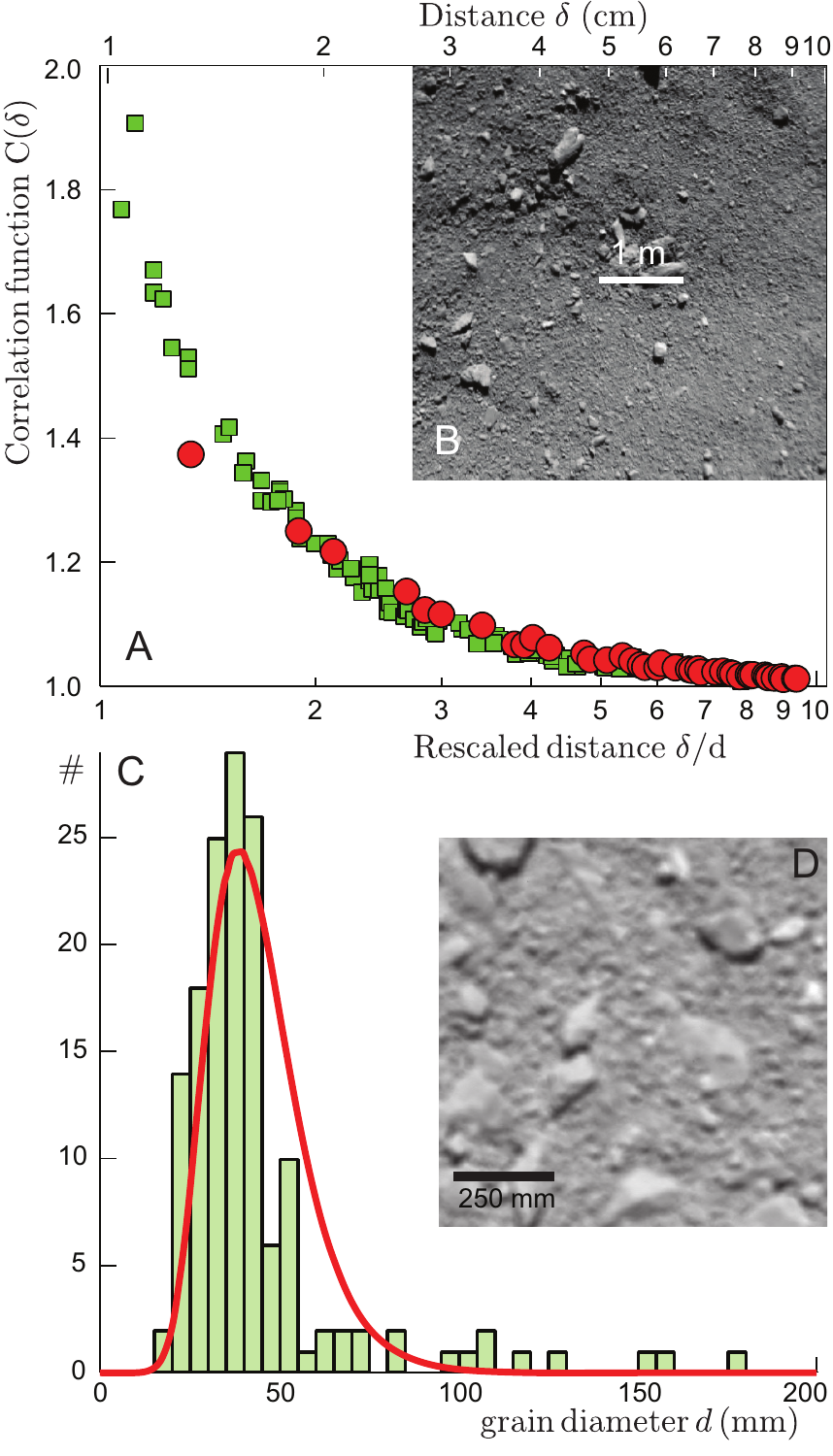}}
\caption{Grain size. \textbf{A} Auto-correlation function $C(\delta)$ (red circles) computed from the photograph of the comet's granular bed, taken by Philae just before its touch down at a site called Agilkia in the Ma'at region (\textbf{B}), where large boulders and rocks have been excluded. The resolution of the picture is 9.5 mm/pixel. Photo credit: ESA/Rosetta/Philae/ROLIS/DLR. The correlation is compared to that computed with pictures of calibrated aeolian sand from the Atlantic Sahara (green square, lower axis, $\delta$ is expressed in units of the grain diameter) taken in the laboratory (Methods). The best collapse of the correlation functions is obtained for a mean grain diameter $d \simeq 9.7$~mm on the comet. $\textbf{C}$ Histogram of grain size $d$ computed from the photograph of the comet's granular bed shown in panel \textbf{D} taken by Rosetta just before its impact in the Ma'at region. The best fit by a log-normal distribution, shown in red, gives a mean grain diameter $d \simeq 38$~mm.}
\label{Fig2}
\end{figure}

Altogether, both seasonal and diurnal time variations of the atmosphere characteristics can be obtained in a simplified spherical geometry (Figs.~3 and S2). At perihelion, we find that the pressure drops by ten orders of magnitude from day to night (Fig.~S2\textbf{b}). The comet's atmosphere therefore presents a strong pressure gradient that drives a tangential flow from the warm, high pressure towards the cold, low pressure regions, in a surface boundary layer (Methods). The extension of the halo of vapor on the dark side of the comet is a signature of this surface wind (Fig.~4B). It reverses direction during the day and is maximal at sunrise and sunset, with a shear velocity $u_*$ on the order of a fraction of the thermal velocity (Fig.~4A). The asymmetry between sunrise and sunset simply results from thermal inertia, as some heat is stored in the superficial layer during the morning and released in the afternoon.

\section{Threshold for grain motion and cohesion}
The vapor density in the coma is still at most seven orders of magnitude lower than that of air on Earth. Can a surface flow with such density and shear velocity entrain grains into motion? The threshold shear velocity $u_t$ above which sediments are transported by a wind is quantitatively determined by the balance between gravity, hydrodynamic drag and cohesive contact force (Methods). Investigating this balance highlights the need to apply findings from contact mechanics of rough interfaces \cite{GT1967} to the study of small solar system bodies. The adhesive free energy, resulting from van der Waals interactions, is proportional to the real area of contact between the grains, which is much smaller than the apparent one because of surface roughness. A realistic computation of this cohesion can be achieved under the assumption that contacts between grains are made of elastically deformed nano-scale asperities and that the apparent area of contact follows Hertz law for two spheres in contact. The cohesive force is then found to scale as the maximal load experienced by the grains to the power $1/3$ (Methods) \cite{RCCBC2002}. Considering that this load is typically the weight of a surface grain, this force scales as $(\rho_p g d/E)^{1/3} \gamma d$, where $\rho_p$ is the grain bulk density, $E$ is the grain Young modulus and $\gamma$ is the surface tension of the grain material. It is therefore much lower than the force $\gamma d$ obtained for ideally smooth grains. Importantly, the gravity force increases as $d^3$, while the cohesive force increases as $d^{4/3}$ only. This allows us to define a cross-over diameter at which these two forces are comparable: $d_m = \left( {\gamma^3}/{E \rho_p^2 g^2} \right)^{1/5}$. It gives the typical grain diameter below which cohesive effects become important and are responsible for the increase of the threshold at small $d$ (Fig.~5). On Earth, this diameter for natural grains is around $10~\mu$m (Fig. S4\textbf{a}). On 67P, making the simple assumption that the values of $E$ and $\gamma$ are similar to those on Earth, the value of $d_m$ can be deduced from the gravity ratio to the power $2/5$: $d_m \simeq (9.8/2.2\,10^{-4})^{2/5} \times 10~\mu {\rm m} \simeq 700~\mu{\rm m}$. Such a millimeter scale is three orders of magnitude smaller than the capillary length $\sqrt{\gamma/\rho_p g}\simeq 1$~m suggested by traditional approaches, which ignore contact roughness \cite{KPMBK2012}.

\begin{figure}[p]
\centerline{\includegraphics{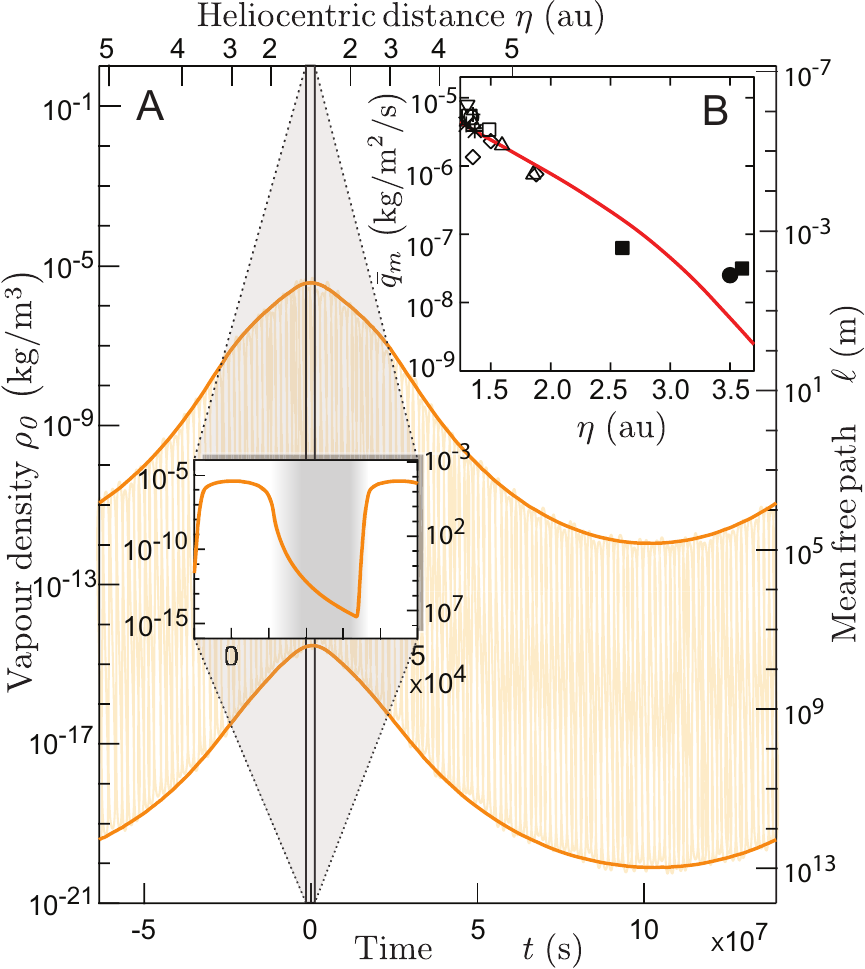}}
\caption{Vapor density and outgassing. \textbf{A} Time evolution of the vapor density $\rho_0$ (left axis) and the corresponding mean free path $\ell \propto 1/\rho_0$ (right axis) just above the comet's surface, calculated along the comet's orbit around the sun in an ideal spherical geometry (Methods). Time is counted with respect to the zenith, at perihelion. Bold orange lines: envelopes of the daily variations (inset), emphasizing the maximum and minimum values. Inset: Zoom on the time evolution of $\rho_0$ and $\ell$ during one comet rotation at perihelion. The  day/night alternation is suggested by the background grey scale. \textbf{B} Global outgassing flux $\bar{q}_m$ as a function of the comet's heliocentric distance $\eta$. Solid line: prediction of the model. Symbols: data from the literature: $\square$ from \cite{S2006}; {\large $\diamond$} from \cite{Hanneretal1985}; $\triangle$ from \cite{OSM1992}; $\ast$, $\triangledown$ and {\large $\circ$} from \cite{BCQS2014} corresponding to data of 2009, 2002 and 1996 respectively; $+$ from \cite{OKKUSOSOS2012}; {\large $\bullet$} from \cite{Gulkisetal2015}; $\blacksquare$ from \cite{B2015}.}
\label{Fig3}
\end{figure}

A second difference with Earth is the large mean free path $\ell$ of the vapor molecules, which leads to a reduced drag force for grains smaller than $\ell$ (Supporting Information). This explains that the threshold velocity $u_t$, plotted as a function of the grain size $d$ (Fig.~5), presents a plateau extending from the millimeter scale to the meter scale (Methods). In conclusion, we find that, sufficiently close to perihelion, all these grains, and in particular those at the centimeter scale observed by Rosetta near bedforms, can be transported by the afternoon thermal wind (Fig.~4). Importantly, this is only a small fraction of the time -- typically $\simeq 6.9 \, 10^3$~s at perihelion, i.e. $\simeq 15 \%$ of the comet's day of $12.4$~h. The asymmetry between sunrise and sunset winds has an important consequence: the morning thermal wind is not strong enough to entrain grains.

\section{Emergent wavelength}
Aeolian dunes and subaqueous ripples form by the same linear instability, which is now well modeled and quantitatively tested against laboratory measurements \cite{CAC2013}. The destabilizing effect results from the phase advance of the wind velocity just above the surface with respect to the elevation profile (Fig.~6B). The stabilizing mechanism comes from the space lag between sediment transport and wind velocity. It is characterized by the saturation length $L_{\rm sat}$, defined as the sediment flux relaxation length towards equilibrium \cite{CAC2013,SKH2001,ACP2010}. As all other parameters are known, $L_{\rm sat}$ is the key quantity selecting the most unstable wavelength $\lambda$. Applying linear stability analysis for 67P (Methods), we compute this wavelength, and empirically find that it approximatively scales as $\lambda \approx L_{\rm sat}^{3/5} \left(\nu/u_*\right)^{2/5}$ (Fig.~6A).

\begin{figure}[p]
\centerline{\includegraphics{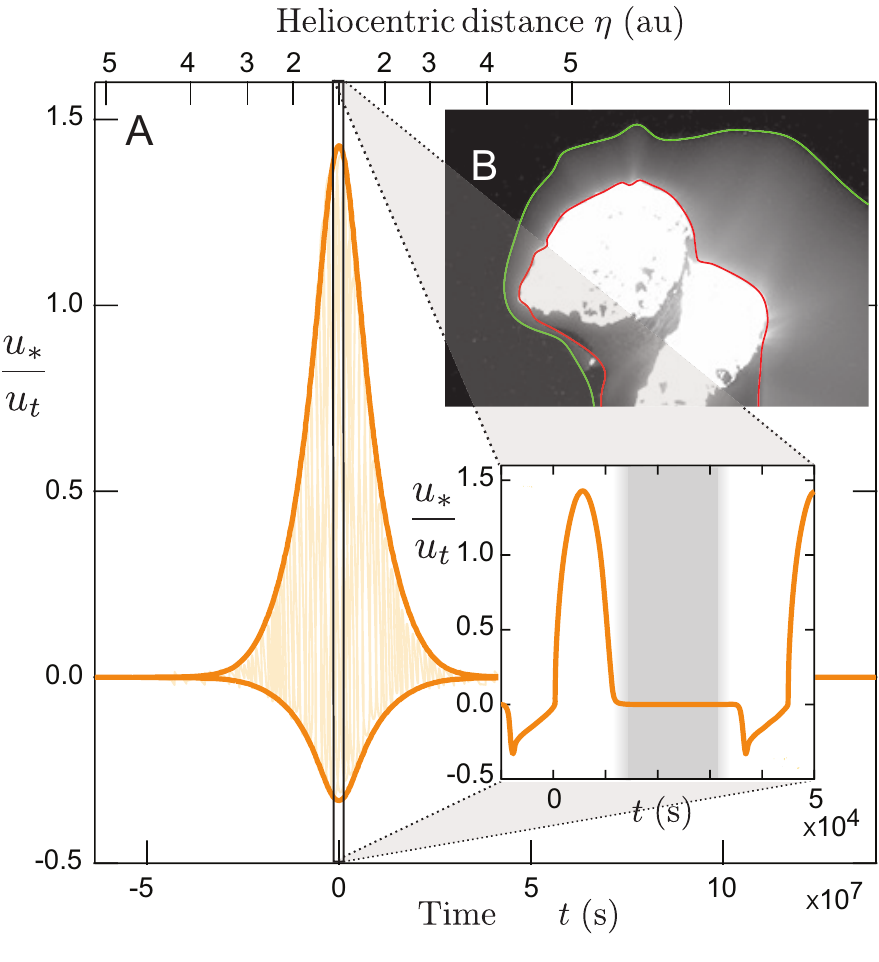}}
\caption{Winds at sunrise and sunset. {\bf A} Time evolution of the velocity ratio $u_*/u_t$, calculated along the comet's orbit around the sun. Time is counted with respect to the zenith, at perihelion. Bold orange lines: envelopes of the daily variations (inset), emphasizing the maximum and minimum values. Inset: Zoom of the evolution of $u_*/u_t$ during one comet day, at perihelion. The day/night alternation is suggested by the background grey scale. Wind is above the transport threshold in the afternoon (counted positive) and in the morning (counted negative). {\bf B} Picture of the comet and its close coma. Red line shows the contour of the comet. Green line shows the contour of the vapor halo at the resolution of the instrument. Some vapor is present on the dark side of the comet even if the vapor sources are located on the illuminated side, providing evidence for the presence of winds. Image taken on 18 February 2016, when Rosetta was 35.6 km from the comet, with a resolution of 3.5 m/pixel. Photo credit: ESA/Rosetta/MPS.}
\label{Fig4}
\end{figure}

With the experience of terrestrial deserts, one can recognize the morphology of new born dunes whose crest-to-crest distance provides a good estimate of $\lambda$: they should be sufficiently young not to present a slip face but sufficiently old to be organized into a regularly spaced pattern. Depending on the location, the crest-to-crest distance is measured in the range $5$--$25$~m (Tab.~S1). Making an analogy with sediment transport processes on larger bodies -- by transposing scaling laws established for saltation --, the analog of aeolian dunes \cite{CAC2013,CA2006,ECA2005} would have an emergent wavelength of $10^8$~m due to the extremely large density ratio on the comet, i.e. much larger than the comet itself. Similarly, using the comet's values, the analogue for aeolian ripples \cite{DCA2014} would produce a pattern of wavelength $10^4$~m. As the other elements (asymmetric shape, granular bed, surface wind above transport threshold) do point to bedforms of the dune family, we conclude that the cometary sediment transport is specific and is associated with a saturation length on the order of $10$~cm.

\section{Sediment transport and bedforms}
Given the very large density ratio $\rho_p/\rho_0$ between grains and vapor, the length needed to accelerate grains to the wind velocity is around $600$\,km for centimeter scale grains. This is much larger than the comet size, meaning that the grains actually keep a velocity $u^p$ negligible in front of the wind velocity $u$. The moving grains are thus submitted to an almost constant drag force equal to that when the grains are static. We then argue that the mode of sediment transport along the comet's surface is traction, where grains remain in contact with the substratum on which they roll or slide. Traction is a slow mode of transport, where the energy brought by the flow is dissipated during the collision of moving grains with the static grains of the bed. Sediment transport on the comet is therefore analogous to subaqueous bedload (Fig. S5). Adapting Bagnold's approach to the comet (Supporting Information), the sediment flux is proportional to the product of the number of moving grains per unit surface and their mean horizontal velocity \cite{DAC2012}. In the subaqueous bedload case, because the density ratio $\rho_p/\rho_0$ is on the order of a few units (in the range $2$--$4$), the moving grains quickly reach a velocity $u^p$ comparable to that of the fluid $u$. On the comet, the constant mechanical forcing resembles, for the thin transport layer, a granular avalanche, in which dissipation comes from the collisions between the grains and is increasing with $u^p$ \cite{A2007}. In that case, close enough to the threshold, the grain velocity follows the scaling law $u^p \sim \sqrt{gd}\simeq 10^{-3}$~m/s and the density of moving grains is a fraction of $1/d^2$, which means that all the grains of this surface transport layer move. The corresponding volume sediment flux $q_{\rm sat}$ therefore scales as $q_{\rm sat} \approx g^{1/2} d^{3/2}$.

Beside the separation of scales between $u^p$ and $u$,  there are important differences with Earth that prevent a cometary saltation  \cite{ThomasetalAA2015} in which the grains would move by bouncing or hopping \cite{OCA2011,KPMBK2012}. The flow is turbulent above a viscous sub-layer, typically $0.7$~m thick at perihelion, where turbulent fluctuations are damped by viscosity. After a rebound, grains with enough energy to reach the turbulent zone would be entrained into suspension, since the settling velocity is much smaller than turbulent fluctuations (Supporting Information). These grains would acquire a vertical velocity larger than the escape velocity, on the order of a meter per second, and would eventually be ejected into the coma.

\begin{figure}[p]
\centerline{\includegraphics{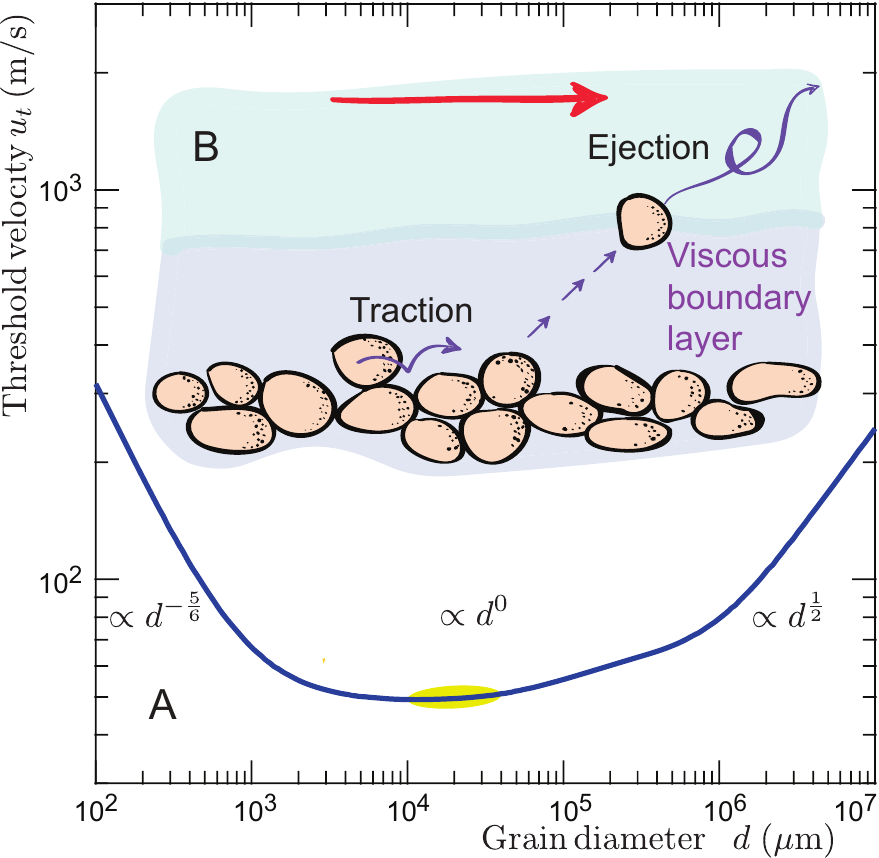}}
\caption{Grain motion. {\bf A} Dependence of the threshold shear velocity $u_t$ with the grain diameter $d$ at perihelion, for afternoon conditions. The minimal velocity above which sediment transport takes place is computed from the force balance on a grain between hydrodynamic drag, bed friction and Van der Waals cohesive forces (Methods). The threshold increases above $d \simeq 1$~m due to gravity and below $d \simeq 1$~mm due to cohesion. In between, $u_t$ is almost constant and on the order of $50$~m/s due to the large mean free path of the vapor $\ell \simeq 3$~cm. Yellow mark: range of observed grain sizes (Fig.~2). {\bf B} Schematic of the vapor flow (red arrow) above the granular bed. Grains rebounding on the bed can reach the upper turbulent zone and are eventually ejected in the coma, which prevents the existence of saltation. The only mode of sediment transport along the bed is traction. Violet background: viscous sub-layer close to the bed, typically $10 \nu/u_* \simeq 0.7$~m thick close to perihelion.}
\label{Fig5}
\end{figure}
\begin{figure}[p]
\centerline{\includegraphics{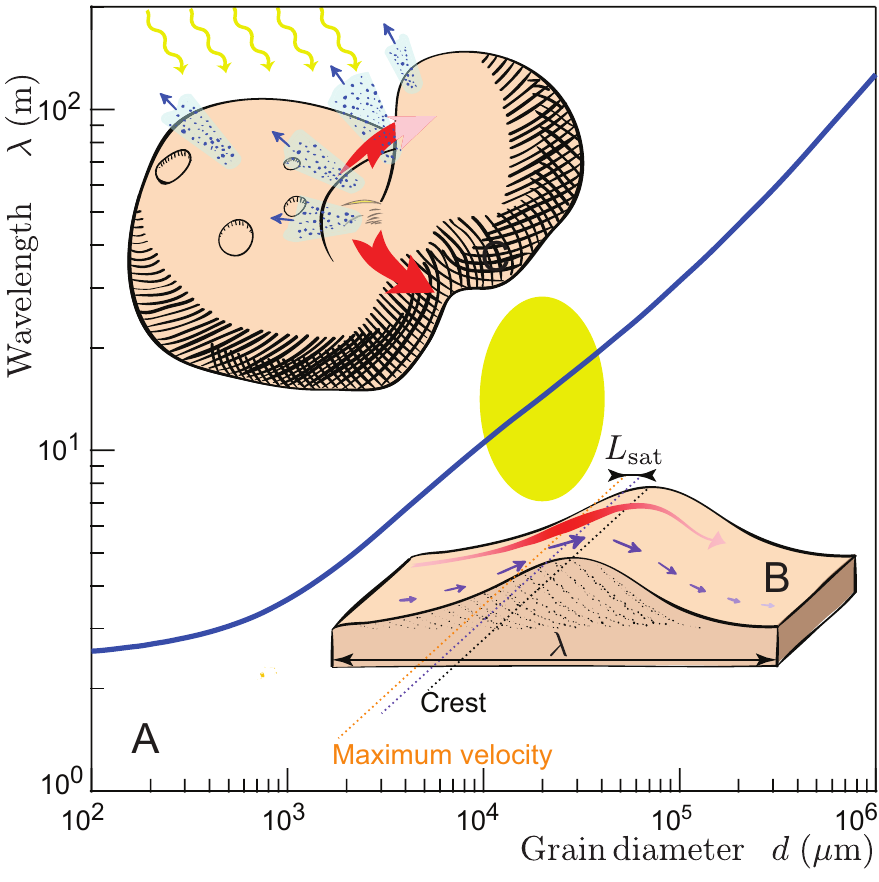}}
\caption{Ripple wavelength. {\bf A} Relation between the wavelength and the mean grain diameter predicted at perihelion, for afternoon conditions. The most unstable mode of the linear instability (Methods) selects the emergent wavelength, which depends on the grain diameter through the saturation length $L_{\rm sat}$ (Fig. S5b). Yellow mark: range of measured crest-to-crest distance and grains size (Tab.~S1). {\bf B} Schematic of the ripple instability mechanism. The wind velocity close to the surface (red arrow) is modulated by the topography and is maximum (red dotted line) upwind of the crest (black dotted line). The sediment flux, which quantifies the amount of transported grains per unit transverse length and unit time, lags behind the wind velocity by the distance $L_{\rm sat}$. Grains are eroded (deposited) when the flux increases (decreases). Instability takes place when the crest is in the deposition zone, i.e. when the maximum of the sediment flux (orange dotted line) is upwind of the crest. {\bf C} Schematic of the outgassing process (blue) and the resulting winds (red arrows) driven by strong pressure gradients from illuminated to shadow areas.}
\label{Fig6}
\end{figure}

We use here the analogy with subaqueous bedload, for which controlled experiments on emerging subaqueous ripples allow us to deduce $L_{\rm sat}/d \simeq 24 \pm 4$ (Fig.~S4\textbf{b}) and retain this law for traction on the comet. As shown in Fig.~6A, for the mean grain diameter $d$ between $10$ and $40$~mm observed in the Ma'at region (Fig.~2), the model predicts an emergent wavelength $\lambda$ between $10$ to $20$~m, in good agreement with the observed crest-to-crest distance (Tab.~S1). For such grains, the traction sediment flux is on the order of $4 \, 10^{-5}$~m$^2$/s. The corresponding ripple growth time, deduced from the linear stability analysis is $\simeq 5 \, 10^4$~s. This time must be compared to the total time during which sediment transport takes place during a revolution around the sun, which is around $10^6$~s ($0.7 \%$ of the revolution period), i.e. 20 times larger. The ripples therefore have enough time to emerge and mature during one comet revolution. In the neck region, pictures of the same location before and after perihelion (Fig.~1B) provide evidence for ripple activity: the smallest ripples have disappeared at the downwind end of the field and a large one has nucleated at the upwind entrance. In between, ripples may have survived and propagated downwind according to the direction of their slip faces. The displacement predicted by the linear stability analysis, on the order of $10$~m, (Fig.~S7) is consistent with the observed pattern shift (Fig.~1B).

\section{Concluding remarks}
We have argued here that the bedforms observed on 67P are likely to be giant ripples, due to their composition, their asymmetric morphology and the existence of surface winds driven by the night/day alternation above the transport threshold. These conclusions are reached from a self-consistent analysis but are of course based on limited data. As bedforms reflect the characteristics of the bed and the flow they originate from, they provide strong constrains of the physical mechanisms at work, which challenge alternative explanations. Comets thus provide an opportunity to better understand erosion and accretion processes on planetesimals, with implications for the open question of how these bodies can grow from the meter to the kilometer scale \cite{Johansenetal2007,JJ2014}.

\vspace*{0.3cm}
\noindent
\rule[0.1cm]{3cm}{1pt}

B. A. is supported by Institut Universitaire de France. P.C. is visiting research associate of the School of Geography of the University of Oxford. P.J. thanks the Natural Science Foundation of China (No. NSFC11402190) for funding. We thank J. Le Bourlot, A.B. Murray, J. Nield and G.S.F. Wiggs for a careful reading of the manuscript.


\newpage
\appendix

\section{Materials and methods}
\label{MM}

We provide here the main ingredients of our analysis and modeling. The Supporting Information gives further technical details on the derivation of the model.

\subsection{Grain size}
Following the technique developed in \cite{CA2006}, a series of calibrated photographs of a sand-bed is used to relate the image auto-correlation to the mean grain diameter $d$ of the bed, whose value is measured independently by sieve analysis. The reference pictures are taken at resolutions going from 1 to 10 pixels per grain diameter. The rescaled correlation functions $C(\delta)$ corresponding to these pictures at different resolutions collapse on a master curve when $\delta$ is divided by $d$ -- both expressed in the same units. To determine an unknown mean grain size from a picture whose resolution is known, one computes its auto-correlation $C(\delta)$, with $\delta$ expressed in meters or in pixels. One then fit by a least square method the value of $d$ that should be used as rescaling factor of $\delta$, to collapse the new curve on the calibration master curve. Even when the grain size is comparable to the resolution, the decay of the correlation between neighboring pixels contains sufficient information to measure $d$ accurately.

\subsection{Ripple propagation}
Two photographs of the same location --~one well before perihelion and the other well after it~-- were used to estimate the bedform propagation distance over one revolution. The photographs are mapped one on the other using fixed elements of relief (cliffs, rocks, holes, etc) that can be recognized on both pictures. The mapping is performed through a projection, assuming in first approximation that the landscape is planar.

\subsection{Thermal balance}
To determine the surface temperature $T_s$ and the vapor mass flux $q_m$ as a function of time (Figs.~3 and S2), we solve the power balance per unit surface
\begin{equation}
(1-\Omega)\psi = \sigma \epsilon T_s^4 + J_s + \mathcal{L} q_m.
\label{Powerbalance}
\end{equation}
This equation relates the solar radiation flux $\psi(t)$ ($\Omega$ is the albedo) to the power radiated according to Stefan's law ($\sigma$ is Stefan's constant and $\epsilon$ the emissivity), to the heat diffusive flux $J_s$ towards the center the nucleus, and to the power absorbed by ice sublimation ($\mathcal{L}$ is the latent heat). Heat diffusion in the nucleus is solved analytically using the decomposition over normal modes in space and time: a mode of frequency $\omega$ penetrates exponentially over a depth $\sqrt{2\kappa_c/|\omega|}$, where $\kappa_c$ is the thermal diffusivity. $J_s$ is therefore related to $T_s$, through a Fourier transform.

\subsection{Porous layer}
To determine the outgassing vapour flux $q_m$, we model the close sub-surface as a thin porous granular layer. Water molecules are emitted from the ice surface located below this porous layer, and make frequent collisions with the grains, in a way analogous to a chaotic billiard. With a probability close to one, they bounce back and are adsorbed again on the ice surface. The probability to cross the porous layer decreases as the inverse of the porous layer thickness $h$. Using the kinetic theory of gasses, the average radial velocity $u_0$ above the layer is determined analytically and corresponds to a Mach number around $0.15$. By contrast, with ice directly in contact with the coma, the outgassing Mach number would have been close to $1$.

\subsection{Turbulent boundary layer}
The pressure gradient along the comet's surface drives a turbulent superficial flow. We model the basal shear velocity $u_*$ associated with this thermal wind, which determines the ability to transport grains along the surface. $u_*$ is related to the surface pressure $p_0$ and to the outgassing velocity $u_0$ by the momentum equation integrated over the thickness of the turbulent boundary layer  $\delta_i$:
\begin{equation}
\rho_0 |u_*| u_*  +  \rho_0 \frac{\Lambda}{\kappa} u_0 u_*= -\frac{\delta_i}{2R}\,\frac{{\rm d} p_0}{{\rm d} \theta},
\label{ustar1}
\end{equation}
where $\Lambda \equiv \ln\left( 1 + \frac{9u_*\delta_i}{\nu} \right)$ is the logarithm of the Reynolds number based on $u_*$, $\delta_i$ and on the viscosity $\nu$. $\delta_i$ is set by the crossover from the inner to the outer layer, i.e. where the inertial terms are comparable to the pressure gradient:
\begin{equation}
\frac{|\Lambda-2|\delta_i}{2\pi\kappa^2 R} \simeq 1  +   \frac{\Lambda}{\kappa} \frac{u_0}{u_*} \, .
\end{equation}
%

\subsection{Cohesion between grains}
The sediment transport threshold depends on the adhesion force $A$ between grains, which is strongly influenced by the grain surface roughness. Considering two grains of diameter $d$ that have been placed in contact by means of a normal load $N$, the apparent area of contact is governed by Hertz law: $a_a \sim (N d/E)^{2/3}$, where $E$ is the Young modulus of the material. However, due to the roughness, the real area of contact $a_r$ is much smaller than the apparent one $a_a$ and, according to Greenwood's theory \cite{GT1967}, is proportional to the normal load: $a_r \sim N/E$. The adhesion force therefore scales as:
\begin{equation}
A \sim \frac{a_a}{a_r}\,\gamma d \sim \gamma \left( \frac{N d}{E} \right)^{1/3}.
\label{CohesiveForce}
\end{equation}
%

\subsection{Sediment transport threshold}
The shear velocity threshold $u_t$ for sediment transport is computed from the force balance applied on a surface grain on the verge to be entrained into motion. Such a grain is submitted to its weight, to a drag force due to the wind flow, to a cohesive force at the grain contacts and to a resistive force associated with the geometrical effect of the surrounding grains. The drag force reads $F_{\rm drag} = \pi/8 \, C_d d^2 \rho_0 u^ 2$, where $u$ is the velocity of the fluid around the grain. The drag coefficient $C_d$ depends on the grain Reynolds number $ud/\nu$ to describe both viscous and turbulent regimes. We also include Cunningham's correction to account for the case of a dilute gas, when the mean free path $\ell$ becomes comparable to the grain size. The grain weight scales as $\rho_p g d^3$ and sets the normal force $N$ in Eq.~\ref{CohesiveForce}, which gives the adhesion force. The resistive force of the bed is modeled as a friction of effective coefficient $\mu$. The expression of $u_t$ can then be derived analytically (Supporting Information) and takes the form:
\begin{equation}
u_t = u_t^0 \left[ 1+ \left( \frac{d_m}{d} \right)^{5/3} \right]^{1/2},
\label{thresholdut}
\end{equation}
where $d_m$ is the cohesive size defined above. In the large $d$ regime, the turbulent drag essentially balances the friction force:
\begin{equation}
u_t \sim \sqrt{(\rho_p/\rho_0)gd} \propto d^{1/2}.
\label{scalingutlarged}
\end{equation}
In the intermediate regime for which $d_m<d<\ell$, the viscous drag balances the friction force:
\begin{equation}
u_t \sim \sqrt{(\rho_p/\rho_0)g\ell} \propto d^{0}.
\label{scalingutintermediated}
\end{equation}
In the small $d$ regime, the viscous drag balances cohesion:
\begin{equation}
u_t \sim \left( \frac{\rho_p g d}{E} \right)^{1/6} \left( \frac{\gamma \ell}{\rho_p g d^3} \right)^{1/2}  \sqrt{(\rho_p/\rho_0)gd} \propto d^{-5/6}.
\label{scalingutsmalld}
\end{equation}
%

\subsection{Linear stability analysis}
The wavelength $\lambda$ at which bedforms emerge can be predicted by the linear stability analysis of a flat sediment bed. The growth rate $\sigma$ and propagation velocity $c$ of a modulated bed is given by:
\begin{equation}
\sigma = \mathcal{Q}k^2 \, \frac{(\mathcal{B} - \mathcal{S})-\mathcal{A} kL_{\rm sat}}{1+(kL_{\rm sat})^2},
\quad
c = \mathcal{Q}k \, \frac{\mathcal{A}  + (\mathcal{B}-\mathcal{S}) kL_{\rm sat}}{1+(kL_{\rm sat})^2}.
\label{GrowthRate_and_PropagationVelocity}
\end{equation}
In these expressions, $k=2\pi/\lambda$ is the bed wavenumber and $\mathcal{Q}$ is the reference sediment flux. $\mathcal{A}$ and $\mathcal{B}$ are the components of the basal shear stress respectively in phase and in quadrature with the elevation profile, which are determined by hydrodynamics (Supporting Information) \cite{CAC2013}. $L_{\rm sat}$ is the saturation length which reflects the space lag of sediment flux in response to a change of wind velocity. $\mathcal{S}$ encodes the fact that the threshold for transport is sensitive to the bed slope with $\mathcal{S} = \frac{1}{\mu} (u_t / u_*)^2$, where $\mu$ is the avalanche slope for the grains considered.

\newpage

\section{Supporting Information}
\label{SI}

In this document, we provide technical details on the derivation of the model we use to describe the vapor outgassing from the nucleus to the comet's coma, the hydrodynamics of the coma, transport law and transport threshold of sediment at the comet's surface, and finally details on the linear stability analysis of the problem that we use to predict the wavelength, growth rate and propagation speed of the emerging bedforms. This technical content is followed by supplementary figures.

\subsection{Geometry and gravity of the comet}
\label{Gravity}

The value of the gravity on the comet is important for the computation of the threshold for sediment transport. Gravity also enters the hydrodynamical equations of the coma. We provide here a derivation to estimate the gravity acceleration in the region of the neck, where the bedforms that we have primarily studied are located. We then define the define the effective radius of the comet, which is used throughout this modeling.

The gravity field on 67P has been studied by [7]. The comet is composed of two lobes related by a thick neck of radius $R_n \simeq 1$~km. The large lobe has dimensions of $4.1 \times 3.2 \times 1.3$ (in km). It can be approximated as a sphere of effective radius $R_l =(4.1 \times 3.3 \times 1.8)^{1/3} /2 \simeq 1.5$~km, leading to a gravity acceleration at the surface $g_l = \mathcal{G} \frac{4\pi}{3} \rho_c R_l \simeq 1.9 \, 10^{-4}$~m/s$^2$, where $\mathcal{G} = 6.67 \, 10^{-11}$~m$^3$kg$^{-1}$s$^{-2}$ is the gravitational constant and $\rho_c \simeq 470$~kg/m$^3$ an estimate of the comet's bulk mass density. Similarly, the small lobe is $2.6 \times 2.3 \times 1.8$ (in km), which gives an effective radius $R_s \simeq 1.1$~km, and a gravity acceleration at the surface $g_s \simeq 1.5 \, 10^{-4}$~m/s$^2$. In the region of the neck, the gravity acceleration is given by
\begin{equation}
g_n = \left[ \left( g_l \sin\theta_l + g_s \sin\theta_s \right)^2 + \left( g_l \cos\theta_l - g_s \cos\theta_s \right)^2 \right]^{1/2},
\label{gravityneck}
\end{equation}
where we have defined the two angles $\tan\theta_l = R_n/R_l$ and $\tan\theta_s = R_n/R_s$. This expression gives $g_n \simeq 2.2 \, 10^{-4}$~m/s$^2$. This value leads to an escape velocity on the order of $\sqrt{g_n R_n} \simeq 0.5$~m/s, which is three orders of magnitude smaller than the thermal velocity $V_{\rm th} \simeq 500$~m/s. As a consequence, the gravity term in the hydrodynamical equations (\ref{Atmo2}) and (\ref{Atmo3}) is on the order of $g R / V_{\rm th}^2 \simeq 10^{-6}$ and is thus negligible.

Despite this two-lobe shape, we will below work in spherical coordinates, simplifying the geometry of the comet to a sphere of effective radius $R_c$. Here we take $R_c \simeq 1.95$~km, corresponding to an equivalent surface $S_c \simeq 47.7$~km$^2$. An equivalent mass ($M_c \simeq 10^{13}$~kg) would have led to a similar value $\simeq 1.7$~km. We denote by $r$ the radial coordinate that originates at the centre of the nucleus, by $\theta$ the ortho-radial (azimuthal) angle, and by $\varphi$ the polar angle. We shall also make use of the distance $z$ to the comet's surface, counted positive downwards. Furthermore, we neglect the effect of the comet's obliquity.

\subsection{Thermo-hydrodynamics of the comet's atmosphere}
\label{Outgassing}

In order to assess sediment transport at the surface of the comet, we need to estimate the vapor density and the vapor flow in the coma. We describe in this section the thermal and ice sublimation processes, taking into account the existence of a porous granular surface layer, as well as the hydrodynamics of the coma.

\subsubsection{Thermal diffusion in the comet's nucleus}
\label{ThermalDiff}
Inside the nucleus, we write the heat conductive flux as $\vec J=-k_c \vec \nabla T$, where $T$ is the temperature field and $k_c$ is the thermal conductivity. Denoting by $C$ the bulk heat capacity of the comet and $\rho_c$ its bulk mass density, the heat conservation equation reads:
\begin{equation}
\rho_c C \partial_t T = k_c \nabla^2T.
\label{ThermalDiffusion}
\end{equation}
All three parameters $k_c$, $C$ and $\rho_c$ are assumed to be homogeneous. Equivalently, a temperature diffusion equation can be written with a thermal diffusivity $\kappa_c=k_c/(\rho_c C)$. The material constituting the bulk of the comet is a mixture of dust and ice, with a rather large porosity $\mathcal{P}$ on the order of $75\%$ [7]. Its effective thermal inertia $I = \sqrt{k_c\rho_c C}$ has been estimated to be in the range $10$--$50$~Jm$^{-2}$K$^{-1}$s$^{-1/2}$ [18]. Taking $\rho_c \simeq 470$~kg/m$^3$ and $C \simeq (1-\mathcal{P}) \times 10^3$~J/kg/K, we obtain $k_c \simeq 10^{-2}$~W/m/K and $\kappa_c \simeq 10^{-7}$~m$^2$/s.

The time evolution of the temperature of the comet's surface $T_s$ can be decomposed in Fourier modes. Diffusion being linear, we can do the reasoning one particular mode of angular frequency $\omega$, written in complex notations as ${\hat T}_s (\omega)$. Assuming that the flux vanishes at infinity (deep inside the bulk of the comet), the solution of the diffusion equation for the temperature field takes the form:
\begin{eqnarray}
{\hat T}(z,\omega)= {\hat T_s}(\omega)\, \text{\rm exp}\left(-(1- i)z \sqrt{\frac{|\omega|}{2\kappa_c}}\right) \qquad \mbox{for} \quad \omega \leq 0,
\label{TemperatureFieldFourierM} \\
{\hat T}(z,\omega)= {\hat T_s}(\omega)\, \text{\rm exp}\left(-(1+ i)z \sqrt{\frac{|\omega|}{2\kappa_c}}\right) \qquad \mbox{for} \quad \omega > 0.
\label{TemperatureFieldFourierP}
\end{eqnarray}
The penetration length $\delta$ is defined:
\begin{equation}
\delta=\sqrt{\frac{2\kappa_c}{|\omega|}} \, .
\label{PenetrationLength}
\end{equation}
The rotation period of the comet is $\Gamma_d=12.4$ hours, or, equivalently, $\omega_d=2\pi/\Gamma_d= 1.4~10^{-4}~{\rm s^{-1}}$. This gives a diurnal penetrating length $\delta_{d} \simeq 4$~cm, which means that a few tens of cm below the surface, the day-night alternation has no influence on the temperature field. Regarding the seasonal variations, the orbital period is $\Gamma_y=6.44$~years, corresponding to a penetrating length $\delta_y \simeq 3$~m. Conversely, one can compute the time scale corresponding to the size of the comet $\delta_h = R_c$, which gives $\Gamma_h \simeq 10^6$~years. This is the time scale required to get a homogeneous temperature $T_a$ across the whole body. It is much smaller than the age of the comet, which is that of the solar system, i.e. about $4.5 \, 10^9$~years.

\subsubsection{Ice sublimation}
\label{Sublimation}
We hypothesize that the vapor outgassing comes from the sublimation of ice just below the surface of the comet. To sublimate ice at a rate corresponding to a vapor mass flux $q_m$ (in kg per second and per unit surface), a power per unit surface $\mathcal{L} q_m$ is absorbed. $\mathcal{L} \simeq 3 \, 10^6$~J/kg is the latent heat of water ice sublimation. The corresponding power balance writes:
\begin{equation}
(1-\Omega)\psi = \sigma \epsilon T_s^4 + J_s + \mathcal{L} q_m,
\label{Powerbalance}
\end{equation}
where $\Omega=0.05$ is the estimated albedo, Stefan's constant is $\sigma = 5.67 \, 10^{-8}$~W/m$^2$/K$^4$ and $\varepsilon \simeq 0.9$ is the estimated emissivity [18]. $\psi$ is the solar radiation flux received by the comet. We write it at latitude $\varphi$ as $\psi = \sin\varphi \, \psi_{\E} (\eta_{\E}/\eta)^2 \phi$, where $\psi_{\E} \simeq 1360$~W/m$^2$ is the radiation flux received from the sun at $\eta_{\E}=1$~astronomical unit (au). $\eta$ is the heliocentric distance of the comet, which is a known function of time along the comet's orbit. $\phi$ encodes the day-night alternation following $\phi (t) = {\rm max} [\cos(2\pi t/\Gamma_d), 0]$. The heat flux, computed at the comet's surface by $J_s = -k_c \left. \partial_z T \right |_{z=0}$, is determined from its Fourier transform $\hat J_{s}$. Using (\ref{TemperatureFieldFourierM},\ref{TemperatureFieldFourierP}), we close it on $\hat T_{s}$ and obtain:
\begin{eqnarray}
\hat J_{s}(\omega)=(1- i)k_c \sqrt{\frac{|\omega|}{2\kappa_c}} \hat T_{s}(\omega) \qquad \mbox{for} \quad \omega \leq 0,
\label{ConductivefluxM}\\
\hat J_{s}(\omega)=(1+ i)k_c \sqrt{\frac{|\omega|}{2\kappa_c}} \hat T_{s}(\omega) \qquad \mbox{for} \quad \omega > 0.
\label{ConductivefluxP}
\end{eqnarray}

The integration of Eq.~\ref{Powerbalance}, coupled to those describing the vapor flow in the atmosphere as well as in the porous surface layer, is used to predict the time variations of the vapor flux $q_m$ at both daily and yearly scales.

\subsubsection{Hydrodynamics and outer vapor flow}
\label{Hydrodynamics}
The vapor flow in the comet's atmosphere is described by the conservation of mass, momentum and energy:
\begin{eqnarray}
\frac{\partial \rho}{\partial t}+\vec \nabla \cdot (\rho \vec u) & = & 0,
\label{Atmo1}\\
\frac{\partial \rho \vec u}{\partial t}+\vec \nabla \cdot (\rho \vec u \vec u) & = & \rho \vec g - \vec \nabla p + \vec \nabla \cdot {\vec{\vec \tau}},
\label{Atmo2}\\
\frac{\partial }{\partial t} \left[\rho\left(\epsilon+\frac{1}{2}u^2\right)\right] & + & \vec \nabla \cdot\left[\rho\left (w+\frac{1}{2}u^2\right)\vec u\right]
\nonumber \\
& = & \rho \vec g \cdot \vec u + \vec \nabla \cdot ({\vec{\vec \tau}} \cdot \vec u) - \vec \nabla \cdot \vec J, \qquad
\label{Atmo3}
\end{eqnarray}
with the mass density $\rho$, the velocity $\vec u$, the pressure $p$, the stress tensor ${\vec{\vec \tau}}$, the specific energy $\epsilon$, the specific enthalpy $w = \epsilon + p/\rho$, the heat flux $\vec J$ and the gravity acceleration $\vec g$. Taking the density weighted time averaging to get so-called Favre averaged Navier Stokes (FANS) equations, the averaged stress tensor can be expressed as the sum of viscous and turbulent contributions:
\begin{equation}
\tau_{ij} = \rho \nu {\dot \gamma}_{ij} + \rho \nu_t \left[ {\dot \gamma}_{ij} - \frac{1}{3} K \delta_{ij} \right],
\label{StressTensorFANS}
\end{equation}
where we have introduced the shear rate ${\dot \gamma}_{ij} = \partial_j u_i + \partial_i u_j - \frac{2}{3} \partial_k u_k \delta_{ij}$. In the ideal gas approximation, the molecular viscosity $\nu$ can be related to the mean free path
\begin{equation}
\ell=\frac{m}{\sqrt{2}\pi d_w^2 \rho} \, ,
\label{defmeanfreepath}
\end{equation}
and to the thermal velocity
\begin{equation}
V_{\rm th} = \sqrt{\frac{8k_BT}{\pi m}} \, ,
\label{defthermalvelocity}
\end{equation}
defined as the mean magnitude of the velocity of the molecules, by
\begin{equation}
\nu = \frac{1}{3} V_{\rm th} \ell.
\label{defmolecularviscosity}
\end{equation}
$k_B=1.38 10^{-23}$ J/K is the Boltzmann constant, $d_w \simeq 0.34$~nm is water molecule size and $m \simeq 3 \, 10^{-26}$~kg is the mass of a water molecule. The turbulent viscosity can be simply modeled by a first order closure $\nu_t =  L^2 | \dot \gamma |$, where $| \dot \gamma |$ is the modulus of the shear rate tensor, and $L$ is the Prandtl mixing length (see e.g. Eq.~\ref{MixingLength} in section 4), involving the phenomenological von K\'arm\'an constant $\kappa \simeq 0.4$. The normal stress components are closed on the velocity field with $K=\chi^2 | \dot \gamma |$, where $\chi \simeq 2.5$ is a second phenomenological constant. Similarly, the averaged heat flux writes:
\begin{equation}
J_i = - \rho \frac{\gamma}{\gamma-1} \left( \frac{\nu}{\rm Pr} + \frac{\nu_t}{{\rm Pr}_t} \right) \partial_i \frac{p}{\rho} \, ,
\label{HeatFluxFANS}
\end{equation}
where $\gamma=4/3$ the adiabatic expansion coefficient of water vapor, and where ${\rm Pr}$ and ${\rm Pr}_t$ are the Prandtl and turbulent Prandtl numbers, both typically on the order of unity for gases. The averaged energy density has also an internal and a turbulent contribution:
\begin{equation}
e = \rho\epsilon=\frac{1}{\gamma-1}p + \frac{1}{2} \nu_t \rho K.
\label{EnergyDensityFANS}
\end{equation}
Finally, the additional term $u_j\tau_{ij}$ complements the enthalpy contribution $\rho w u_i$. Note also that Coriolis forces have been neglected, as the Rossby number $V_{\rm th}/(R_c \omega_d) \simeq 10^3$ is large.

Eqs. (\ref{Atmo1}-\ref{Atmo3}) can be solved averaging over the polar angle, and assuming steady state. We describe the atmosphere as a two-layer flow: an outer layer where viscosity and turbulent fluctuations can be neglected (perfect flow) and an inner turbulent layer of thickness $\delta_i \ll R_c$ matching with the surface conditions. We separately note $U_r$ and $U_\theta$ the velocity components in outer layer, and $u_r$ and $u_\theta$ those in the inner layer (see next section). This hydrodynamical description of the comet's atmosphere loses it validity when the mean free path of the vapor becomes on the order of the comet size itself.

Neglecting all dissipative terms in (\ref{Atmo1}-\ref{Atmo3}), the steady equations for the outer layer are, for mass conservation:
\begin{equation}
\frac{1}{r^2} \frac{\partial}{\partial r} \left( r^2 \rho U_r \right) + \frac{1}{r} \frac{\partial}{\partial \theta} \left( \rho U_\theta \right) = 0;
\label{Atmo1SphericalMass}
\end{equation}
for momentum conservation in the radial direction:
\begin{equation}
\frac{1}{r^2} \frac{\partial}{\partial r} \left( r^2 \rho U_r^2 \right) + \frac{1}{r} \frac{\partial}{\partial \theta} \left(  \rho U_r U_\theta \right) - \frac{1}{r} \rho U_\theta^ 2 + \frac{\partial p}{\partial r}= 0;
\label{Atmo2SphericalMomentumr}
\end{equation}
for momentum conservation in the ortho-radial direction:
\begin{equation}
\frac{1}{r^3} \frac{\partial}{\partial r} \left( r^3 \rho U_r U_\theta \right) + \frac{1}{r} \frac{\partial}{\partial \theta} \left( \rho U_\theta^2 \right) + \frac{1}{r} \frac{\partial p}{\partial \theta}= 0;
\label{Atmo2SphericalMomentumtheta}
\end{equation}
and for the energy conservation:
\begin{eqnarray}
\frac{1}{r^2} \frac{\partial}{\partial r} \left[ r^2 \left( \frac{1}{2}\rho \left( U_r^2+U_\theta^2 \right) + \frac{\gamma}{\gamma-1}p \right) U_r \right] & + &
\nonumber \\
\frac{1}{r} \frac{\partial}{\partial \theta} \left[ \left( \frac{1}{2}\rho \left( U_r^2+U_\theta^2 \right) + \frac{\gamma}{\gamma-1}p \right) U_\theta \right] & = & 0.
\label{Atmo3SphericalEnergy}
\end{eqnarray}
The asymptotic analysis of these equations gives $U_r\propto r^0$, $U_\theta \propto r^{2(1-\gamma)}$, $\rho\propto r^{-2}$ and $p\propto r^{-2 \gamma}$. One concludes that orthoradial terms are subdominant in the outer layer, so that the equations, at the leading order reduce to
\begin{eqnarray}
\frac{1}{r^2} \frac{\partial}{\partial r} \left( r^2 \rho U_r \right) & = & 0,
\label{Atmo1SphericalMassRadial} \\
\frac{1}{r^2} \frac{\partial}{\partial r} \left( r^2 \rho U_r^2 \right) + \frac{\partial p}{\partial r} & = & 0,
\label{Atmo2SphericalMomentumrRadial} \\
\frac{1}{r^2} \frac{\partial}{\partial r} \left[ r^2 \left( \frac{1}{2}\rho U_r^2 + \frac{\gamma}{\gamma-1}p \right) U_r \right] & = & 0.
\label{Atmo3SphericalEnergyRadial}
\end{eqnarray}
These equations can be analytically integrated as:
\begin{eqnarray}
U_r & = & U_0 \sqrt{G(r)},
\label{uradial} \\
\rho & = & \rho_0 \left( \frac{R_c}{r} \right)^2 \frac{1}{\sqrt{G(r)}},
\label{rhoradial} \\
p & = & \left[ p_0 + \frac{\gamma-1}{2\gamma} \rho_0 U_0^2 \left[ 1 - G(r) \right] \right] \left( \frac{R_c}{r} \right)^2 \frac{1}{\sqrt{G(r)}}, \qquad \\
&=& p_0\left[\frac{G_\infty-G(r)}{G_\infty-1} \right] \left( \frac{R_c}{r} \right)^2 \frac{1}{\sqrt{G(r)}}
\label{pradial}
\end{eqnarray}
where the function $G$ satisfies $G(R_c)=1$, so that  $\rho_0$ and $p_0$ are the vapor density and the pressure at the surface of the comet $r=R_c$ and $U_0$ the vapor velocity at top of the surface layer. We have introduced 
\begin{equation}
G_\infty = 1 + \frac{2\gamma}{\gamma-1} \frac{p_0}{\rho_0 U_0^2}.
\label{defGinfinity}
\end{equation}
From (\ref{Atmo2SphericalMomentumrRadial}), we see that $G$ must satisfy
\begin{equation}
G' - \frac{\gamma-1}{\gamma+1} \left( G_\infty \frac{G'}{G} + (G_\infty-G) \frac{4}{r} \right) = 0,
\label{gradial}
\end{equation}
This first order differential equation solves into:
\begin{equation}
G^{\frac{1}{2}(\gamma-1)} \left( \frac{G_\infty-G}{G_\infty-1} \right) = \left( \frac{R_c}{r} \right)^{2 (\gamma-1)} .
\label{gradial2}
\end{equation}
The outer radial vapor flow is then entirely determined by the three surface quantities $\rho_0$, $U_0$ and $p_0$.

\subsubsection{Turbulent boundary layer}
\label{TurbulentBoundaryLayer}
We need to computed the vapor wind flow close to the surface, which may entrain the surface grains into motion. This flow is controlled by the momentum balance in the boundary layer approximation, in which the horizontal diffusion of momentum is negligible:
\begin{equation}
\frac{1}{r^3} \frac{\partial}{\partial r} \left[ r^3 (\rho u_r u_\theta - \tau_{r \theta}) \right] + \frac{1}{r} \frac{\partial}{\partial \theta} \left(\rho u_\theta^2 \right) + \frac{1}{r} \frac{\partial p}{\partial \theta}= 0.
\label{Atmo2SphericalMomentumthetaInner}
\end{equation}
To compute an approximate solution, we write the velocity profile in the inner layer under the form:
\begin{equation}
u_\theta(r) = \frac{u_*}{\kappa} \, \ln\left( 1 + \frac{r-R_c}{z_0} \right),
\label{mathcalU}
\end{equation}
parametrized by the shear velocity $u_*$ defined from the basal shear stress $\tau_{r \theta}^0 \equiv \rho_0 |u_*| u_*$. For the sake of simplicity, we use here the logarithmic law of the wall, but more complicated profiles could be easily accommodated. $z_0$ is the aerodynamic roughness and here we take $z_0 = 0.11 \nu/u_*$ corresponding to the smooth aerodynamic regime. We introduce the notation
\begin{equation}
\Lambda \equiv \ln\left( 1 + \frac{\delta_i}{z_0} \right),
\label{defLambda}
\end{equation}
where $\delta_i$ is the thickness of the boundary layer.

Estimating the terms in the momentum equation projected along the radial direction, we find that the variation $\delta p$ of pressure across the boundary layer scales like $\delta p \sim (\delta_i/R_c)^2 p_0 $, which is small compared to $p_0$. From the energy equation, the scaling of the temperature variation across the boundary layer is similarly: $\delta T \sim \delta_i/R_c T_0$. The pressure, temperature and density in the inner layer can thus be considered as constant (with respect to $r$): $p \simeq p_0$, $T \simeq T_0$ and $\rho \simeq \rho_0$.

The radial velocity at the top of the of the boundary layer is $U_0$. Integrating (\ref{Atmo2SphericalMomentumthetaInner}) between $r=R_c$ and $r=R_c+\delta_i$, for $\delta_i \ll R_c$, we obtain:
\begin{eqnarray}
\rho_0 |u_*| u_* + \rho_0 \frac{\Lambda}{\kappa} U_0 u_* & + &  \frac{\rm d}{{\rm d} \theta} \left[ \frac{\left(2 - 2\Lambda +\Lambda^2\right) \delta_i}{\kappa^2 R_c} \rho_0 u_*^2 \right]
\nonumber \\
& + & \frac{\delta_i}{R_c}\,\frac{{\rm d} p_0}{{\rm d} \theta} = 0,
\label{ustar1}
\end{eqnarray}
where we have used the fact that the velocity $u_\theta$ vanishes at the comet's surface, and that the shear stress vanishes at the top of the inner turbulent boundary layer, when one reaches the outer perfect flow.

The radial component of the velocity in the inner layer $u_r$ is deduced from $u_\theta$ by the mass conservation equation:
\begin{equation}
\frac{1}{r^2} \frac{\partial}{\partial r} \left( r^2 \rho u_r \right) + \frac{1}{r} \frac{\partial}{\partial \theta} \left( \rho u_\theta \right) = 0.
\label{Atmo1SphericalMomentumthetaInner}
\end{equation}
By integration across the boundary layer, we similarly obtain:
\begin{equation}
U_0 = u_0 - \frac{1}{ \rho_0} \frac{\rm d}{{\rm d} \theta} \left[ \frac{(\Lambda -1)\delta_i}{\kappa R_c} \rho_0 u_* \right].
\label{ustar2}
\end{equation}
Using this expression for $U_0$ in (\ref{ustar1}), we deduce:
\begin{eqnarray}
\rho_0 |u_*| u_* & + & \rho_0 \frac{\Lambda}{\kappa} u_0 u_*
\nonumber \\
& + & \frac{\rm d}{{\rm d} \theta} \left[ \frac{\left(2 - 2\Lambda +\Lambda^2\right) \delta_i}{\kappa^2 R_c} \rho_0 u_*^2 \right]
\nonumber \\
& - & \frac{\Lambda}{\kappa} u_* \frac{\rm d}{{\rm d} \theta} \left[ \frac{(\Lambda -1)\delta_i}{\kappa R_c} \rho_0 u_* \right]
= -\frac{\delta_i}{R_c}\,\frac{{\rm d} p_0}{{\rm d} \theta}, \qquad
\label{ustar3}
\end{eqnarray}

The boundary layer thickness corresponds to the crossover altitude at which one makes the transition from the inner to the outer layer, i.e. where the inertial terms are comparable to the pressure gradient:
\begin{eqnarray}
\left(|u_*| + \frac{\Lambda}{\kappa} u_0 \right) \rho_0 u_* & \approx & -\frac{\rm d}{{\rm d} \theta} \left[ \frac{\left(2 - 2\Lambda +\Lambda^2\right) \delta_i}{\kappa^2 R_c} \rho_0 u_*^2 \right]
\nonumber \\
& + & \frac{\Lambda}{\kappa} u_* \frac{\rm d}{{\rm d} \theta} \left[ \frac{(\Lambda -1)\delta_i}{\kappa R_c} \rho_0 u_* \right],
\label{ustar5}
\end{eqnarray}
so that (\ref{ustar3}) simplifies into:
\begin{equation}
-\frac{\delta_i}{2R_c}\,\frac{{\rm d} p_0}{{\rm d} \theta} = \left(|u_*| + \frac{\Lambda}{\kappa} u_0 \right) \rho_0 u_*.
\label{ustar4}
\end{equation}
Eq.~\ref{ustar5} is further simplified under the assumption that variations of all quantities along $\theta$ are slow, essentially equivalent to sinusoidal variations, i.e. with $\frac{{\rm d}}{{\rm d} \theta} \approx \frac{1}{2\pi}$. We then obtain:
\begin{equation}
\frac{|\Lambda-2|\delta_i}{2\pi\kappa^2 R_c} \approx 1  +   \frac{\Lambda}{\kappa} \frac{u_0}{u_*}.
\label{ustar6}
\end{equation}
For given density $\rho_0(\theta)$ and pressure $p_0(\theta)$ profiles, we finally solve (\ref{ustar4}) and (\ref{ustar6}) to obtain $u_*$ as well as $\delta_i$. Note that the above equations are only valid if the thickness of the turbulent boundary layer is larger than that of the viscous sub-layer, i.e. when $\delta_i \gtrsim 10 \nu/u_*$.

\subsubsection{Porous sub-surface layer}
\label{PorousLayer}
We describe the close sub-surface as a thin porous granular layer of thickness $h$. The picture is that of a chaotic billiard, where a water molecule, emitted at depth $z=h$ where the ice is, experiences collisions with the grains of the packing but not with the other molecules. The mean free path of the molecules is then a fraction of grain size $d$. The probability for a molecule to cross this layer rather than going back to $z=h$ and being adsorbed by the ice again is $p_c \propto d/h$, depending on porosity and grain shape.

We assume that the water molecules emitted from ice have a half Maxwell-Boltzmann velocity distribution:
\begin{equation}
P_i(\vec v) = \left( \frac{m}{2\pi k_B T_i} \right)^{3/2} \exp \left( - \frac{m |\vec v|^2}{2 k_B T_i} \right)\Theta(\vec v\cdot \vec e_r).
\label{distriPs}
\end{equation}
$T_i$ is the temperature of the ice at $z=h$. $\Theta$ is the Heaviside function and $\vec e_r$ is the unit vector pointing upwards. The vapor mass flux of molecules emitted by the ice surface is then
\begin{equation}
F \rho_{\rm sat} \int_{-\infty}^{+\infty} \!\!\! dv_x \int_{-\infty}^{+\infty} \!\!\! dv_y \int_0^{+\infty} \!\!\! dv_r \,\, v_r P_i(\vec v) = \frac{1}{4} F \rho_{\rm sat} V_{\rm th}^i,
\label{defqm}
\end{equation}
where $F$ is the ice surface fraction, and where we have introduced the thermal velocity $V_{\rm th}^i = V_{\rm th}(T_i) = \sqrt{8k_B T_i/(\pi m)}$ (see Eq.~\ref{defthermalvelocity}). $\rho_{\rm sat}$ is the saturated vapor density, here also evaluated at the temperature of the ice $T_i$.

At the comet's surface ($z=0$), where the temperature of the vapor is $T_0$, we assume furthermore that the vapor flow has an average velocity $u_0 \vec e_r$, so that the water molecules have a velocity distribution given by:
\begin{equation}
P_0(\vec v) = \left( \frac{m}{2\pi k_B T_0} \right)^{3/2} \exp \left( - \frac{m |\vec v-u_0\vec e_r|^2}{2 k_B T_0} \right).
\label{distriP0}
\end{equation}
The vapor mass flux of molecules entering in the porous layer from the atmosphere, whose density is $\rho_0$, is then
\begin{eqnarray}
q_- & = & \rho_0 \int_{-\infty}^{+\infty} \!\!\! dv_x \int_{-\infty}^{+\infty} \!\!\! dv_y \int_{-\infty}^{0} \!\!\! dv_r \,\, (-v_r) P_0(\vec v)
\nonumber \\
& = & \frac{1}{4} f(\Upsilon_0) \rho_0 V_{\rm th}^0,
\label{defqminus}
\end{eqnarray}
where we have introduced the thermal velocity $V_{\rm th}^0 = V_{\rm th}(T_0)$, the velocity ratio $\Upsilon_0 \equiv u_0/V_{\rm th}^0$ and defined the function:
\begin{equation}
f(\Upsilon)= e^{-\frac{4\Upsilon^2}{\pi}}-2\Upsilon \left[1-\text{erf}\left(\frac{2\Upsilon}{\sqrt{\pi} }\right)\right].
\label{deffunctionf}
\end{equation}
$\Upsilon$ is similar to a Mach number, as the speed of sound in an ideal gas is $\sqrt{\gamma k_B T/m} = \sqrt{\pi/6} \, V_{\rm th}$ for an adiabatic index $\gamma=4/3$ used here.

Assuming perfect absorption of the water molecules when they come back to ice (a vanishing probability of rebound), the vapor mass flux coming out at the surface $q_m=\rho_0 u_0$ is then the result of the following balance:
\begin{equation}
\Upsilon_0 \rho_0 V_{\rm th}^0=p_c \left(\frac{F}{4} \rho_{\rm sat} V^i_{\rm th}-q_-\right).
\label{MassFluxSurface}
\end{equation}
In the limit of an unlimited ($F=1$) and vanishingly thin ($T_i=T_0$) layer, the Hertz-Knudsen sublimation law, with a vapor flux proportional to $(\rho_{\rm sat} - \rho_0)V_{\rm th}$ is recovered. Similarly, the momentum flux $\rho_0 u_0^2 + p_0$ reads:
\begin{equation}
\left(\Upsilon_0^2  + \frac{\pi}{8} \right) \rho_0 {V_{\rm th}^0}^2= \frac{\pi}{4}\left[\frac 14  F p_c \rho_{\rm sat} {V^i_{\rm th}}^2+(2-p_c)q_- V_{\rm th}^0 \right]. \qquad
\label{MomentumFluxSurface}
\end{equation}
Finally, the energy flux $\left( \frac{1}{2} \rho_0 u_0^2 + \frac{\gamma}{\gamma-1} p_0 \right) u_0$ reads:
\begin{equation}
\frac{1}{2}\Upsilon_0\left(\Upsilon_0^2 + \pi \right) \rho_0 {V_{\rm th}^0}^3 = \frac{7\pi}{16} p_c \left[ \frac 14  F \rho_{\rm sat} {V^i_{\rm th}}^3-q_- {V_{\rm th}^0}^2  \right].
\label{EnergyFluxSurface}
\end{equation}
Introducing the expression for $q_-$ (\ref{defqminus}) into Eqs. \ref{MassFluxSurface} and \ref{MomentumFluxSurface}, we solve for $\rho_0$ and $V^0_{\rm th}$:
\begin{eqnarray}
\rho_0 & = & F p_c \, \frac{\pi [f(\Upsilon_0)(p_c-2)+2]+16 \Upsilon_0^2}{\pi [f(\Upsilon_0) p_c+4 \Upsilon_0]^2} \, \rho_{\rm sat},
\label{rho0solved} \\
V^0_{\rm th} & = & \frac{\pi [f(\Upsilon_0) p_c+4 \Upsilon_0]}{\pi [f(\Upsilon_0) (p_c-2)+2]+16 \Upsilon_0^2} \, V^i_{\rm th}.
\label{Vth0solved}
\end{eqnarray}
The final equation for $\Upsilon_0$ is obtained introducing these expressions into (\ref{EnergyFluxSurface}):
\begin{eqnarray}
- 7\pi^2+\left(32 \pi^2 - 112\pi \right) \Upsilon_0^2 + \left(32 \pi - 448 \right) \Upsilon_0^4 & + &
\nonumber \\
7(p_c-1) f^2(\Upsilon_0) + \left[(14-7p_c)\pi^2+15 p_c \pi^2 \Upsilon_0 \right. & + &
\nonumber \\
\left. (112\pi-56 p_c \pi) \Upsilon_0^2+8p_c\pi \Upsilon_0^3\right] f(\Upsilon_0)  & = & 0. \qquad 
\label{FinalUpsilon0}
\end{eqnarray}

To solve numerically this equation, values must be chosen for the different parameters. Consistently with the value of the porosity of the comet's ground, we take $F=0.2$ for the ice surface fraction. The porous layer thickness is set to $h=1.5 \, d$, which corresponds to a mono-layer of grains not attached to the icy bed, and free to move by the wind. The probability for a water molecule to cross the porous layer is set to $p_c = 0.1 d/h \simeq 0.07$, in order to adjust the vapor density at the comet's surface (see below). With these numbers, the velocity ratio $\Upsilon_0 = u_0/\sqrt{8k_B T_0/(\pi m)}$, which compares the outgassing velocity to the thermal velocity of the vapour at the comet's surface can be computed as the solution of Eq.~\ref{FinalUpsilon0}. Its value is remarkably insensitive to $p_c$, and is always around $\Upsilon_0=0.11$, corresponding to a Mach number $\simeq 0.15$.

\subsubsection{Global vapor flux}
\label{vaporRate}
Observations [18,19,21,22,23] provide data for the global outgassing flux of the comet at different heliocentric distances $\eta$ (Fig.~3B), which we use to calibrate some parameters of the model. From the local vapor mass flux $q_m$ coming out at the surface, integrated over the whole comet, the global vapor flux reads:
\begin{equation}
\bar{q}_m(\eta) = \frac{\alpha}{4\pi} \int_{-\pi}^{\pi} \!\! d\theta \int_0^{\pi} \!\! \sin\varphi \, q_m(\theta,\varphi) \, d\varphi,
\label{Rotationvaporrate1}
\end{equation}
where the factor $\alpha$ accounts for the fraction of the surface where sublimation is effective. Assuming that all points of the surface receiving the same insolation would produce the same vapor rate, one can solve Eq.~\ref{Powerbalance} at the equator only ($\varphi=\pi/2$) and compute the vapor rate as
\begin{equation}
\bar{q}_m(\eta) = \frac{\alpha}{4} \int_{-\pi}^{\pi} \!\! |\sin\theta| \, q_m(\theta) \, d\theta,
\label{Rotationvaporrate2}
\end{equation}
where the angle $\theta=0$ points in the direction of the sun. This assumption is valid as long as the heat flux term $J_s$ in (\ref{Powerbalance}) is negligible, so that the surface points can be considered as thermally decoupled. This is the case in the illuminated side of the comet ($-\pi/2 \le \theta \le \pi/2$), where most of the vapor flux comes from. This approximation is uncontrolled on the night-side, where $J_s$, due to the thermal inertia of the comet's body, is the source of heat for sublimation, but corresponding to a negligible part of $\bar{q}_m$. 

The fit of the observational data allows us to set the porous layer thickness to $h=1.5 \, d$. Larger values lead to a dependence of the vapor flux $\bar{q}_m$ that decreases too fast with the heliocentric distance $\eta$. Also, the fraction of active (sublimating) surface is adjusted to $\alpha=0.1$ in order to reproduce the value of the flux at perihelion.

\subsection{Sediment transport}
\label{Transport}

For given vapor density and flow, we need to know whether the wind is able to set the surface grains into motion. We first compute the threshold for transport and then derive the transport law, accounting in both cases for the peculiar conditions of the comet's atmosphere.

\subsubsection{Transport threshold}
\label{Threshold}
We consider a grain of size $d$ at the surface of the comet, on the verge to be entrained into motion. It is submitted to its weight, to a cohesive force at the grain contacts and to a resistive force associated with the geometrical effect of the surrounding grains. The later can be modeled by a Coulomb friction of coefficient $\mu$ relating the tangential and normal forces. The grain weight can be expressed as $\frac{\pi}{6} \rho_p g d^3$, where $g$ is the gravity acceleration and $\rho_p$ is the mass density of the grains. The relevant dimensionless parameter to quantify the ability of the fluid to put the grains of the bed into motion is the Shields number defined as $\Theta = \tau/[(\rho_p-\rho)gd]$, where $\rho$ is the fluid density and $\tau$ is the shear stress exerted by the fluid on the bed.

In most practical cases, the threshold velocity falls in the cross-over between the viscous and turbulent asymptotic regimes. It is thus important to have a model of it valid in both regimes [9]. The drag force exerted on a grain reads
\begin{equation}
F_{\rm drag}=\frac{\pi}{8} C_d d^2 \rho u^2,
\label{defFdrag}
\end{equation}
where $u$ is the velocity of the fluid around the grain and $C_d$ is a drag coefficient. In order to account for viscous as well as turbulent regimes, $C_d$ can conveniently be written as:
\begin{equation}
C_d = \left(C_\infty^{1/2}+s \left(\frac{\nu}{ud}\right)^{1/2} \right)^2,
\label{dragcoeff}
\end{equation}
where $\nu$ is the fluid viscosity. $C_\infty$ and $s$ are phenomenological calibrated constants. For example, we have $C_\infty \simeq 1$ and $s \simeq 5$ for natural grains. In the case of dilute gas, i.e. when the mean free path $\ell$ becomes comparable to the grain size, an empirical correction due to Cunningham [36] is applied, and we take
\begin{equation}
s^2=\frac{25}{1+\frac{2\ell}{d}(1.257+0.4 \exp(-0.55d/\ell))} \, .
\label{CunninghamFactor}
\end{equation}

When the grain is at rest at the surface of the bed, we consider that the hydrodynamical stress is exerted on its upper half so that the effective drag force becomes $F_{\rm drag}= \beta \frac{\pi}{8} C_d d^2 u^2$, with $\beta=1/2$. Just at the threshold and neglecting cohesion for the moment, this force is balanced by the horizontal bed friction felt by the grain: $F_t=\frac{\pi}{6} \mu (\rho_p -\rho) g d^3$. Here we take $\mu = \tan(29^{\circ}) \simeq 0.55$. We introduce the viscous size
\begin{equation}
d_\nu = (\rho_p/\rho -1)^{-1/3} \nu^{2/3}g^{-1/3},
\label{defdnu}
\end{equation}
and further make the fluid velocity dimensionless as $\mathcal{S}^{1/2} \equiv u/\sqrt{(\rho_p/\rho-1)gd}$. With these notations, the threshold value of the flow velocity at the scale of the grain, denoted as $\mathcal{S}^{1/2}_t$, is solution of
\begin{equation}
\left( C_\infty \mathcal{S}_t \right )^{1/2} + s \left(\frac{d_\nu }{d}\right)^{3/4} \mathcal{S}_t^{1/4} - \left(\frac{4\mu}{3\beta}\right)^{1/2} = 0,
\label{balancedragfriction}
\end{equation}
which resolves immediately into:
\begin{eqnarray}
\mathcal{S}_t=\frac{1}{16 C_\infty^2}\left[\left( s^2\left(\frac{d_\nu }{d}\right)^{3/2} \right. \right. \!\! & + & \left. \left. 8 \left(\frac{\mu C_\infty}{3\beta}\right)^{1/2}\right)^{1/2} \right.
\nonumber \\
& - & \left. s  \left(\frac{d_\nu }{d}\right)^{3/4}\right]^4.
\label{St}
\end{eqnarray}
Following [9], the corresponding threshold Shields number is the sum of a viscous and a turbulent contribution:
\begin{equation}
\Theta_t = 2 \left( \frac{d_\nu}{d} \right)^{3/2} \!\! \mathcal{S}_t^{1/2} + \frac{\kappa^2}{\ln^2(1+1/2\xi)} \, \mathcal{S}_t,
\label{Thetat}
\end{equation}
where $\xi$ is the hydrodynamic roughness rescaled by the grain diameter. Here we take the experimental value $\xi=1/30$.

The grains of the bed also feel an adhesion force $A$ that results from van der Waals interactions. This force depends on the real surface of the grains in contact, and therefore on the normal force exerted on the grains. A realistic computation of this cohesion can be achieved under the assumption that contacts between grains are made of many nano-scale asperities. Whether these micro-contacts are in an elastic or in a plastic state, the resulting scaling laws are essentially the same, and $A$ can be expressed as
\begin{equation}
A \propto \left( \frac{\rho_p g d}{E} \right)^{1/3} \gamma d,
\label{Nadhesion}
\end{equation}
where $E$ is the grain Young modulus and $\gamma$ is the surface tension of the grain material [9]. While the gravity force increases as $d^3$ the cohesive force increases as $d^{4/3}$ only. The cross-over diameter $d_m$ at which these two forces are comparable is:
\begin{equation}
d_m = \left( \frac{\gamma^3}{E (\rho_p g)^2} \right)^{1/5} \!\! .
\label{dm}
\end{equation}
$d_m$ gives the typical grain diameter below which cohesive effects become important and are responsible for the increase of the threshold at small $d$. For silica (quartz) grains on Earth, the cohesive size $d_m$ is around $10$~$\mu$m, and this is why sand grains, with typical diameters on the order of a few hundreds microns, are not affected by cohesion. On the comet, the composition of the regolith dust is not precisely know, but the particle bulk density $\rho_p$ has been estimated in the range $1000$--$3000$~kg/m$^3$ [37], i.e. close to that of sand on Earth. We make the assumption that the values of $E$ and $\gamma$ are also similar for the particles on both bodies. According to (\ref{dm}), the ratio of the $d_m$ values on Earth and 67P is then essentially given by the corresponding ratio of the gravities, to the power $2/5$. Using the gravity field derived above, we can estimate $d_m \simeq (9.8/0.00022)^{2/5} \times 10~\mu {\rm m} \simeq 720~\mu{\rm m}$ on the comet. Accounting for these cohesion effects, the threshold Shields number finally reads:
\begin{equation}
\Theta_t = \Theta_t^0 \left[ 1 + \frac{3}{2} \left( \frac{d_m}{d} \right)^{5/3} \right],
\label{Thetatcohesion}
\end{equation}
where $\Theta_t^0$ is the expression given by Eq. \ref{Thetat} [9]. This expression is used to plot $u_t \equiv \sqrt{\Theta_t (\rho_p/\rho-1)gd}$ as a function of $d$ in Fig.~5, and shows a minimum value on the order of $50$~m/s for the whole range $10^3$--$10^5$~$\mu$m.

A similar approach can be used to compute the settling velocity $V_{\rm fall}$, which also gives the vertical threshold velocity, balancing the drag force and the particle weight. We can proceed as in Eq.~\ref{balancedragfriction}, but with $\mu/\beta=1$ and get:
\begin{eqnarray}
\mathcal{S}_{\rm fall}=\frac{1}{16 C_\infty^2}\left[\left( s^2\left(\frac{d_\nu }{d}\right)^{3/2} \right. \right. \!\! & + & \left. \left. 8 \left(\frac{C_\infty}{3}\right)^{1/2}\right)^{1/2} \right.
\nonumber \\
& - & \left. s  \left(\frac{d_\nu }{d}\right)^{3/4}\right]^4.
\label{Sfall}
\end{eqnarray}
The settling velocity is found always smaller than $u_*$ during the fraction of time when sediment transport occurs.

Another effect is electric charging of the grains. As recently reviewed in [40], the literature reports surface electrification on the order of $2 \cdot 10^{-2}$~C/m$^2$, which originates from separating two contacting surfaces. This can induce grain electric charging when a contact between two grains opens. Hertz contact law $(\rho_p g d^4/E)^{2/3} \simeq 5 \cdot 10^{-12}$~m$^2$ provides an upper bound of the contact area -- we consider grains of density $\rho_p \simeq 2 \cdot 10^3$~kg/m$^3$, of diameter $d \simeq 10^{-2}$~m and of Young modulus $E \simeq 50$~GPa, and with a gravity acceleration $g \simeq 2 \cdot 10^{-4}$~m/s$^2$ at the comet's surface. This gives a charge per grain $e \simeq 10^{-13}$~C/grain, which is consistent with other reported grain electrification values [40]. The corresponding electric force can be estimated as $\frac{1}{4\pi\epsilon_0} \left( \frac{e}{d} \right)^2 \simeq 10^{-12}$~N, where $\epsilon_0=8.85 \cdot 10^{-12}$~F/m is the vacuum permittivity. This force is to be compared to the weight of the grain $\frac{\pi}{6} \rho_p g d^3 \simeq 2 \cdot 10^{-5}$~N. Under this first order assumption, we can then neglect the effect of grain electrification in the computation of the transport threshold as well as in the estimate of the sediment flux.

\subsubsection{Saturated transport flux}
\label{SaturatedTransportFlux}
The grains on the comet's bed move in the traction mode. We derive here the corresponding sediment flux at saturation $q_{\rm sat}$, i.e. in the steady and homogeneous case. The saturated flux can generally be expressed as:
\begin{equation}
q_{\rm sat}=\frac{1}{\phi_b} \frac{\pi}{6} d^3 N u^p,
\label{SaturatedFlux}
\end{equation}
where $\phi_b$ is the bed volume fraction, $N$ is the number of moving grains per unit surface and $u^p$ is their mean horizontal velocity [34]. $q_{\rm sat}$, in m$^2$/s, counts the volume of the grains (packed at the bed volume fraction) passing a vertical surface of unit width, and per unit time.

In the sub-aqueous bed-load case, because the density ratio $\rho_p/\rho$ is on the order of a few units, the drag length is equal to a few $d$. The moving grains then quickly reach a velocity comparable that of the fluid $u$. In the cometary case, however, this drag length is much larger than the comet size, so that $u^p$ remains much smaller than $u$. This gives an almost constant drag force $F_{\rm drag}$ on the moving grains, equal to that when the grains are static. This situation of constant mechanical forcing then resembles, for the thin transport layer, a granular avalanche, in which dissipation comes from the collisions between the grains and is thus increasing with $u^p$ [35]. In that case, it has been shown that, close enough to the threshold, the grain velocity follows the scaling law $u^p \sim \sqrt{gd}$, with a multiplicative factor around unity [38].

From Bagnold's original idea, the basal shear stress $\tau = \rho u_*^2$ is decomposed into the sum of the grain-borne and fluid-borne contributions $\tau^p + \tau^f$. The grain-borne stress is $\tau^p=N F_{\rm drag}$. The fluid-borne stress must be the threshold stress $\tau_t = \rho u_t^2$ at equilibrium transport. We then obtain $N= (\tau-\tau_t)/F_{\rm drag}$.  Combined, these expressions give:
\begin{equation}
q_{\rm sat} \sim \frac{1}{\phi_b} \frac{\pi}{6} d^3 \frac{\tau-\tau_t}{F_{\rm drag}} \sqrt{gd}.
\label{SaturatedFlux1}
\end{equation}
For $\tau$ on the order of a few $\tau_t$, the number of moving grains per unit surface soon reaches $N \simeq 1/d^2$, which means that all the grains of this surface transport layer move, leading to a typical flux on the order of
\begin{equation}
q_{\rm sat} \approx g^{1/2} d^{3/2}.
\label{ApproxSaturatedFlux}
\end{equation}
From the two measurements in the Ma'at region, we estimate a typical grain size $d=2$~cm. The flux is then around $q_{\rm sat} \simeq 4 \,10^{-5}$~m$^2$/s. In the neck (Hapi) region, the observed wavelength suggest a smaller grain size on the order of $d \simeq 4$~mm, which gives $q_{\rm sat} \simeq 4 \,10^{-6}$~m$^2$/s.

\subsection{Linear stability analysis}
\label{LSA}

The linear stability analysis gives the time and length scales at which bedforms emerge from a flat bed [4]. We use it to calibrate the model in the bedload case, with experimental data on sub-aqueous ripples. We also use it to predict the characteristics of the bedforms on the comet.

\subsubsection{Dispersion relation}
\label{DispRel}
The growth rate $\sigma$ and propagation velocity $c$ of a bed modulation of the wavenumber $k=2\pi/\lambda$, where $\lambda$ is the wavelength, is given by:
\begin{eqnarray}
\sigma & = & \mathcal{Q}k^2 \, \frac{(\mathcal{B} - \mathcal{S})-\mathcal{A} kL_{\rm sat}}{1+(kL_{\rm sat})^2},
\label{GrowthRate} \\
c & = & \mathcal{Q}k \, \frac{\mathcal{A}  + (\mathcal{B}-\mathcal{S}) kL_{\rm sat}}{1+(kL_{\rm sat})^2}.
\label{PropagationVelocity}
\end{eqnarray}
In these expressions, $\mathcal{Q} \equiv \tau \partial_\tau q_{\rm sat}$ quantifies the sediment transport, and, in the cometary case, we take for its value the above scaling law (\ref{ApproxSaturatedFlux}): $\mathcal{Q} \approx g^{1/2} d^{3/2}$. $\mathcal{A}$ and $\mathcal{B}$ are the components of the basal shear stress in phase and in quadrature with the bottom, respectively (see below). $\mathcal{S}$ encodes the fact that the threshold for transport is sensitive to the bed slope with $\mathcal{S} = \frac{1}{\mu} \tau_t/\tau$, where $\mu = \tan(29^{\circ}) \simeq 0.55$ is the tangent of the avalanche angle. The saturation length $L_{\rm sat}$ gives the length scale over which sediment transport relaxes towards equilibrium. Comparing the prediction of this analysis to experimental measurements of the wavelength of emerging sub-aqueous ripples, we can calibrate the behavior of the saturation length in the bedload case. We obtain $L_{\rm sat}/d \simeq 24$, independent of the velocity of the flow (Fig.~S4).

\subsubsection{Basal shear stress on an undulated bed}
\label{AandB}
The shear stress exerted by a flow in the $x$-direction on a fixed granular bed of elevation $z=Z(x)$ can be computed by means of hydrodynamic equations. Here we use Reynolds averaged Navier Stokes equations:
\begin{eqnarray}
\partial_i u_i  & = & 0,
\label{NS1}\\
\rho \partial_t u_i + \rho u_j \partial_j u_i  & = & \partial_j \tau_{ij}-\partial_i p,
\label{NS2}
\end{eqnarray}
where $p$ is the pressure, $\tau_{ij}$ contains the Reynolds stress tensor and is closed on the velocity field $u_i$ with a Prandtl-like first order turbulence closure as:
\begin{equation}
\tau_{ij}/\rho =  \left ( L^2 |\dot \gamma| + \nu \right ) \dot \gamma_{ij}- \frac{1}{3} \chi^2 L^2 |\dot \gamma|^2 \delta_{ij}.
\label{Closure1stOrder}
\end{equation}
In this expression, $\nu$ is the fluid viscosity, $|\dot \gamma| = \sqrt{\frac{1}{2} \dot \gamma_{ij} \dot \gamma_{ij}}$ is the strain rate modulus, where we have introduced the strain rate tensor $\dot \gamma_{ij}=\partial_i u_j+\partial_j u_i$, and $\chi$ is a phenomenological constant typically in the range $2$--$3$. $L$ is the mixing length, for which we adopt a van Driest-like expression:
\begin{equation}
L = \kappa (z+r_L d-Z) \left[1-\exp\left(-\frac{\tau_{xz}^{1/2}(z+s_L d-Z)}{\nu R_t}\right)\right],
\label{MixingLength}
\end{equation}
where $\kappa \simeq 0.4$ is the von K\'arm\'an constant, $d$ the grain size, $r_L=1/30$ and $s_L=1/3$ are dimensionless numbers and $R_t$ is a transitional Reynolds number. Following [4,39], $R_t$ depends on a dimensionless number ${\mathcal H}$ which depends on, but lags behind, the pressure gradient:
\begin{equation}
\alpha_H \frac{\nu}{u_*}\partial_x {\mathcal H}=\frac{\nu}{u_*^3} \partial_x (\tau_{xx}-p)-{\mathcal H}
\label{Hanra1}
\end{equation}
where $\alpha_H\simeq 2000$ is the multiplicative factor in front of the space lag. We also introduce $\beta_H\simeq 35$ as the relative variation of $R_t$ due to the pressure gradient:
\begin{equation}
\beta_H=\frac{1}{R_t^0} \;\frac{d R_t}{d {\mathcal H}}>0,
\label{Hanra2}
\end{equation}
where the transitional Reynolds number for the homogeneous case is $R_t^0=25$.

When the bed is modulated as $Z(x) = \zeta e^{ikx}$, these equations can be linearized with respect to the small parameter $k\zeta$ and solved for non-slip conditions on the bed and vanishing first order corrections at $z \to \infty$. The shear stress takes the generic form
\begin{equation}
\tau_{xz} =  \rho u_*^2 \left[1+k\zeta e^{ikx} S_t\right],
\label{Hanra3}
\end{equation}
where $S_t$ is a dimensionless function of the rescaled vertical coordinate $kz$. $\mathcal{A}$ and $\mathcal{B}$ are defined as $S_t(0) = \mathcal{A} + i \mathcal{B}$. They are functions of $k\nu/u_*$, as displayed in Fig.~S6.

\subsubsection{Most unstable mode}
\label{MostUnstable}
A most unstable mode $k_m$ corresponding to the maximum growth rate is deduced from (\ref{GrowthRate}) as the solution of $\frac{d\sigma}{dk}=0$. The corresponding growth rate is $\sigma_m = \sigma(k_m)$, the growth time is $1/\sigma_m$, and the propagation speed is $c_m = c(k_m)$. This mode is displayed as a red dot on the curves showing the dispersion relations (Fig.~S7).

\newpage
\centerline{\includegraphics{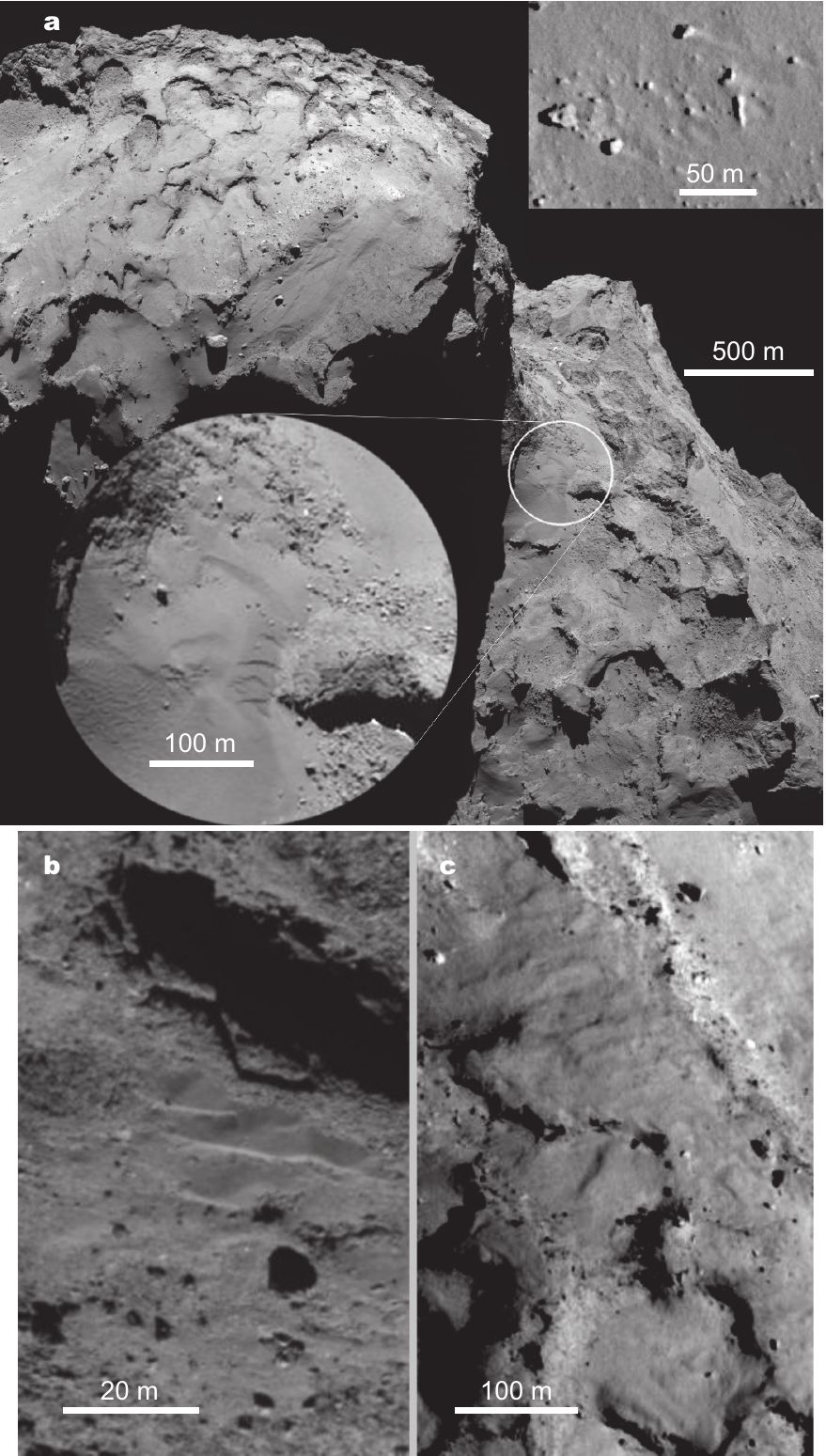}}
\noindent
{{\bf Figure S1.}
\textbf{a} Photographs of the ripples in the neck `Hapi' region, from which the yellow marks in Fig.1 were deduced. Image taken on 17 January 2016, when Rosetta was 85.2 km from Comet 67P/Churyumov-Gerasimenko, with a resolution of 1.55 m/pixel. Upper right inset: photographs of wind tails (or shadow dunes) behind boulders. Image taken on 18 September 2014.  \textbf{b} Photograph of ripples in `Maftet' region. Image taken on 05 March 2016, when Rosetta was 20.3 km from  67P, with a resolution of 0.36 m/pixel. \textbf{c} Photograph of ripples at `Hatmehit' region. Image taken on 13 April 2016, when Rosetta was 109.2 km from 67P, with a resolution of 1.98 m/pixel. All Photo credits: ESA/Rosetta/MPS (see also Tab.~S1).
}

\newpage
\centerline{\includegraphics{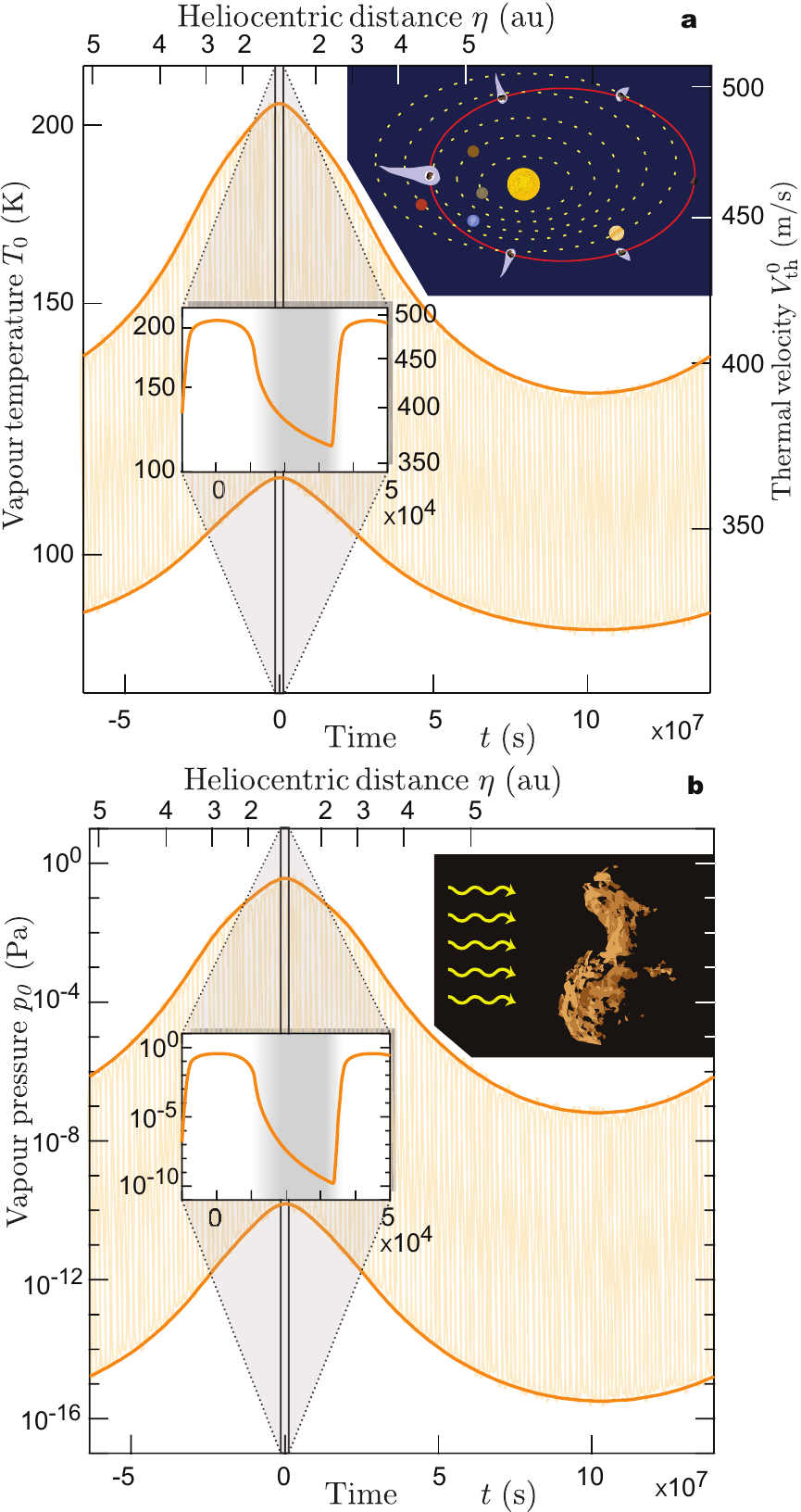}}
\noindent
{{\bf Figure S2.}
Computation of the vapor characteristics at the surface of the comet. The model (detailed in the Suplementary Material) is based on a thermal balance between different effects. The comet receives energy from the sun and radiates some energy back to space with a power related to the ground surface temperature by Stefan's law. Thermal inertia leads to a storage/release of internal energy over a penetration depth which is meter scale for seasonal variations and centimeter scale for daily variations. Ice sublimation requires an input of energy equal to the product of the vapor flux and the latent heat.  This process occurs in the close sub-surface, and the net outgassing vapor flux involves a balance between emission of water molecules at the base of the porous surface granular layer (Fig. S3) and absorption of molecules proportionally to the surface atmosphere density. Both seasonal and diurnal time variations of the atmosphere characteristics are computed in an ideal spherical geometry. \textbf{a} Time evolution of the vapor temperature $T_0$ (left axis) and corresponding thermal velocity $V_{\rm th}^0 \propto \sqrt{T_0}$ (right axis) just above the comet's surface, calculated along the comet's orbit around the sun (inset schematics). Time is counted with respect to the zenith, at perihelion. Bold orange lines: envelopes of the daily variations (inset), emphasizing the maximum and minimum values. Inset: Zoom on the time evolution of $T_0$ and $V_{\rm th}^0$ during one comet rotation at perihelion. The  day/night alternation is suggested by the background grey scale. \textbf{b} Same for the vapor pressure $p_0$. Top inset in panel a: sketch of the comet's trajectory around the sun. Top inset in panel b: photo of the comet illuminated by the sun, as suggested by the wavy yellow arrows.
}

\newpage
\centerline{\includegraphics{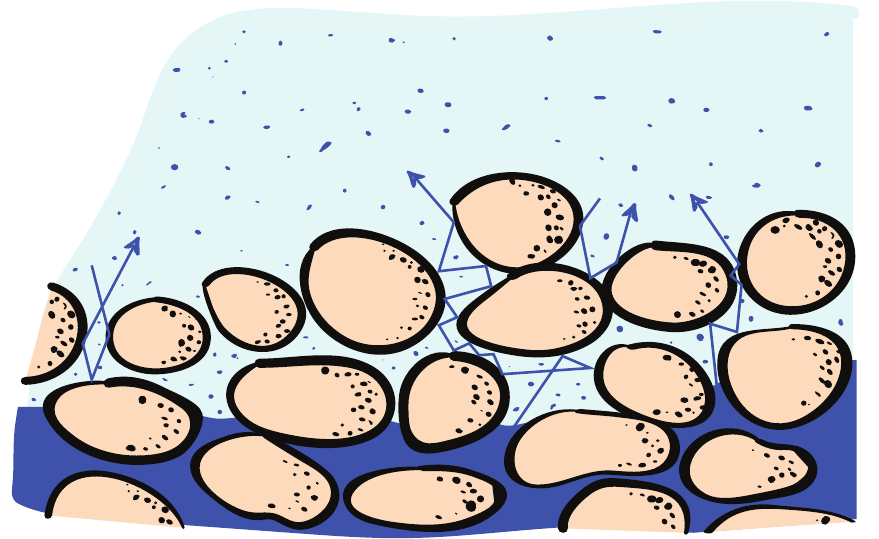}}
\noindent
{{\bf Figure S3.}
Schematics of the porous granular layer at the comet's surface. The water molecules are emitted by the ice (dark blue) at the thermal velocity corresponding to the ice temperature. Experiencing collisions with the grains of the packing (blue arrows), the molecules have a probability to cross the layer decreasing as the inverse of its thickness. The mean free path of the molecules in the layer is comparable to the pores between the grains, i.e. a fraction of the grain size. Molecules just above the surface may also enter the porous layer and be absorbed if they reach the ice. This layer is typically $1.5 d$ thick, so that the surface grains, not glued to ice, are potentially free to move if the wind is above the transport threshold.
}

\newpage
\centerline{\includegraphics{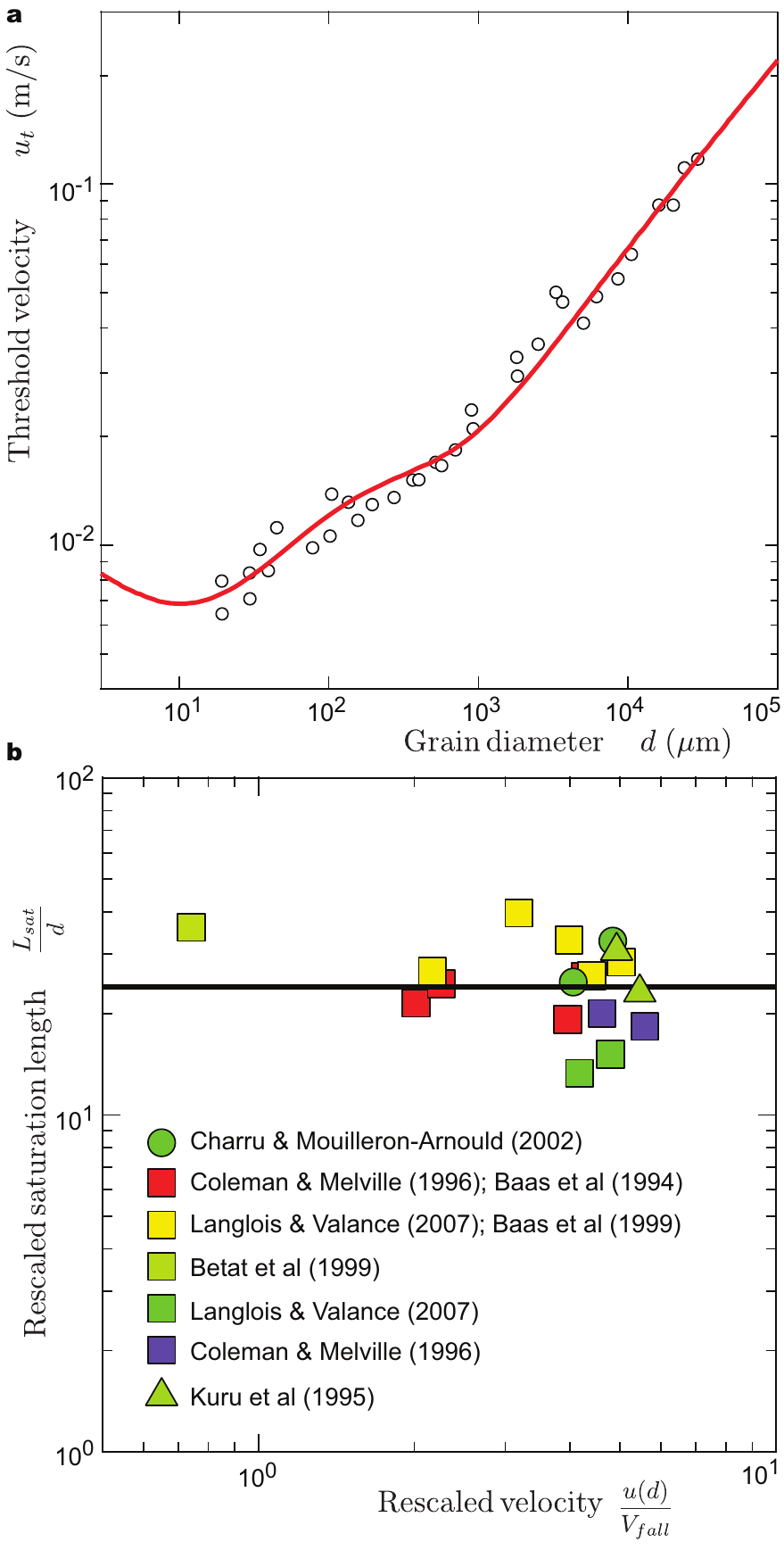}}
\noindent
{{\bf Figure S4.}
Calibration of the model under water, using data obtained in laboratory experiments. \textbf{a} Dependence of the threshold shear velocity $u_t$ on the grain diameter $d$. The best fit of experimental measurements (symbols) by theoretical predictions gives a cohesive diameter $d_m \simeq 10$~microns. Data from Yalin \& Karahan, Hydraul. Div., Am. Soc. Civil Eng \textbf{105}, 1433 (1979). \textbf{b}  Saturation length $L_{\rm sat}$ in units of $d$ as a function of the flow velocity at a grain size above the surface $u(d)$ rescaled by the grain settling velocity $V_{\rm fall}$. $L_{\rm sat}$ is deduced from the measurement of the wavelength of emerging ripples, corresponding to the fastest growing mode. Data are obtained for various experimental conditions: grains in oil (circles), in water (squares), and in water-glycerin solution (triangles); color codes for the grain size from 100 microns (red) to 830 microns (violet). Black solid line: $L_{\rm sat}/d \simeq 24$.
}

\newpage
\centerline{\includegraphics{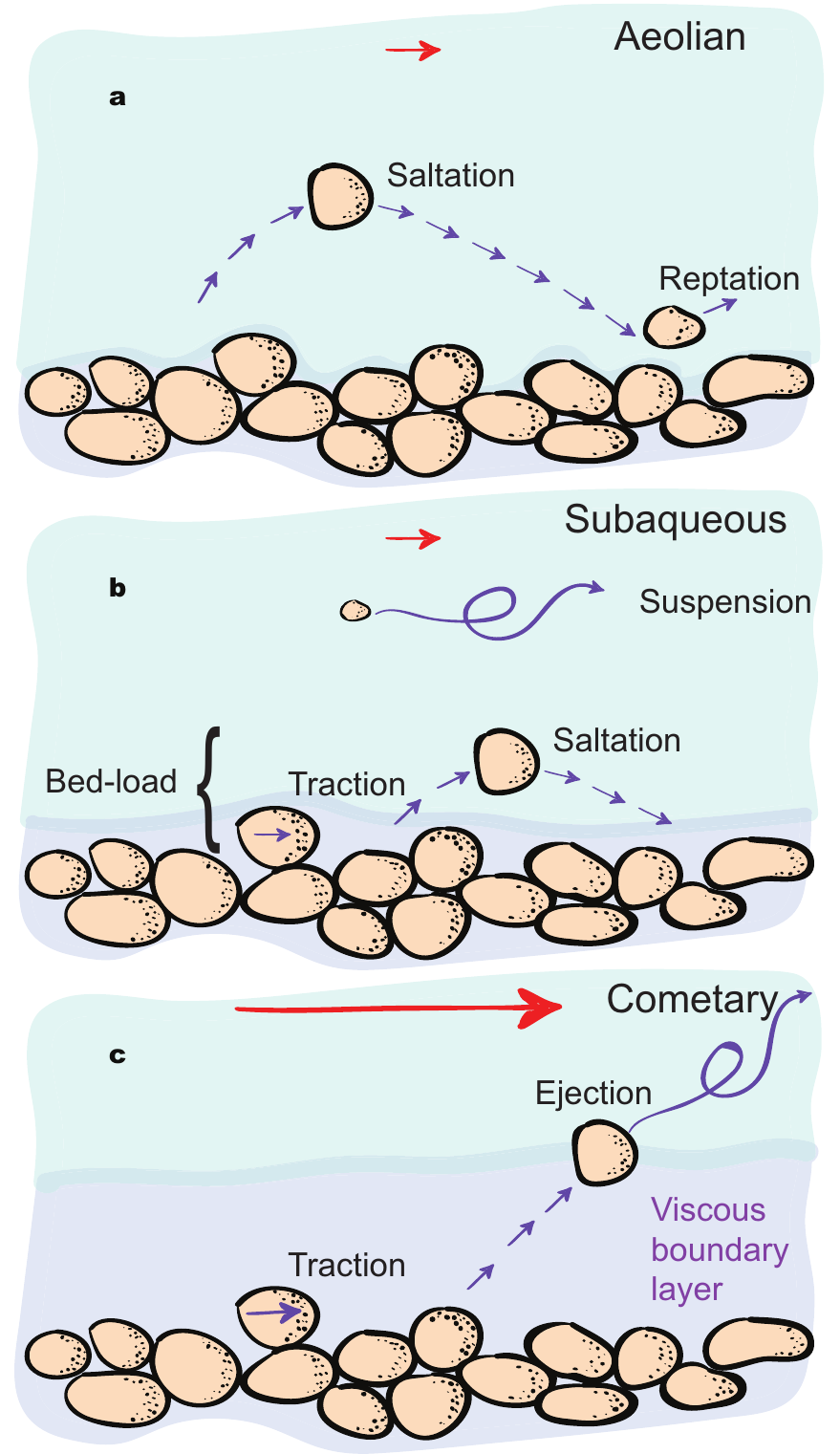}}
\noindent
{{\bf Figure S5.}
Schematics featuring the modes of sediment transport in the aeolian, subaqueous and cometary cases. \textbf{a} In the aeolian case, the density ratio $\rho_p/\rho$ is large so that the grains are mainly transported in saltation, in a succession of jumps. When the impact of saltating grains on the bed is strong enough, they release a splash-like shower of ejected grains that make small hops, and this secondary transport mode is called reptation. \textbf{b} In the subaqueous case, the grains and the fluid have comparable densities. The transport is mainly a turbulent suspension when the velocity of turbulent fluctuations is larger than the settling velocity. When gravity is large enough to confine sediment transport in a layer at the surface of the bed, one refers to bedload: the grains are either hopping in saltation or roll and slide at the bed surface, with long-contacts between the grains (traction). \textbf{c}  In the cometary case, grains rebounding on the bed are eventually ejected in the coma, which prevents the existence of saltation. The only mode of sediment transport along the bed is traction. This schematics holds for monolithic (crystalline) grains as well as for agglomerates of smaller particles. Violet background: viscous sub-layer close to the bed, which is typically $10 \nu/u_* \simeq 0.7$~m thick in the cometary case at perihelion.
}

\newpage
\centerline{\includegraphics{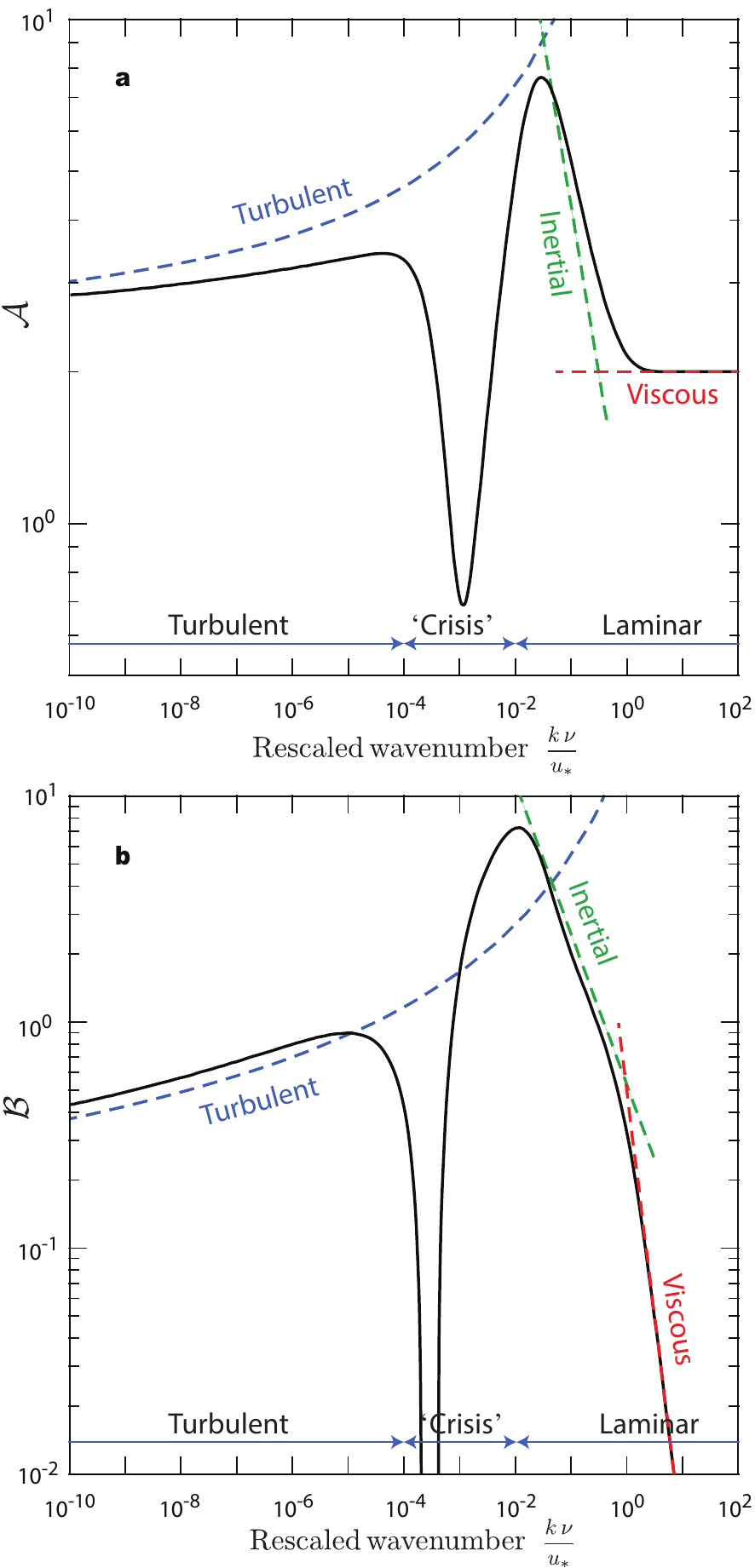}}
\noindent
{{\bf Figure S6.}
Basal shear stress components $\mathcal{A}$ in phase (\textbf{a}) and $\mathcal{B}$ in quadrature (\textbf{b}) with respect to the bed elevation, as functions of the rescaled wave number $k\nu/u_*$. This quantity is the inverse of the Reynolds number based on the wavelength, which can be interpreted as a Reynolds number for the perturbation. Depending on $k\nu/u_*$, three asymptotic regimes can be identified, where the disturbed pressure gradient is balanced by the turbulent Reynolds stress (blue dashed line), by inertia (green dashed line) and by the viscous stress (red dashed line) respectively. The laminar regime is separated from the turbulent regime by a transitional region where a `crisis' can be observed. The principle of the computation of $\mathcal{A}$ and $\mathcal{B}$ is explained in the Supportive Information, see also [4] for discussion and comparison of such curves with experimental data.
}

\newpage
\centerline{\includegraphics{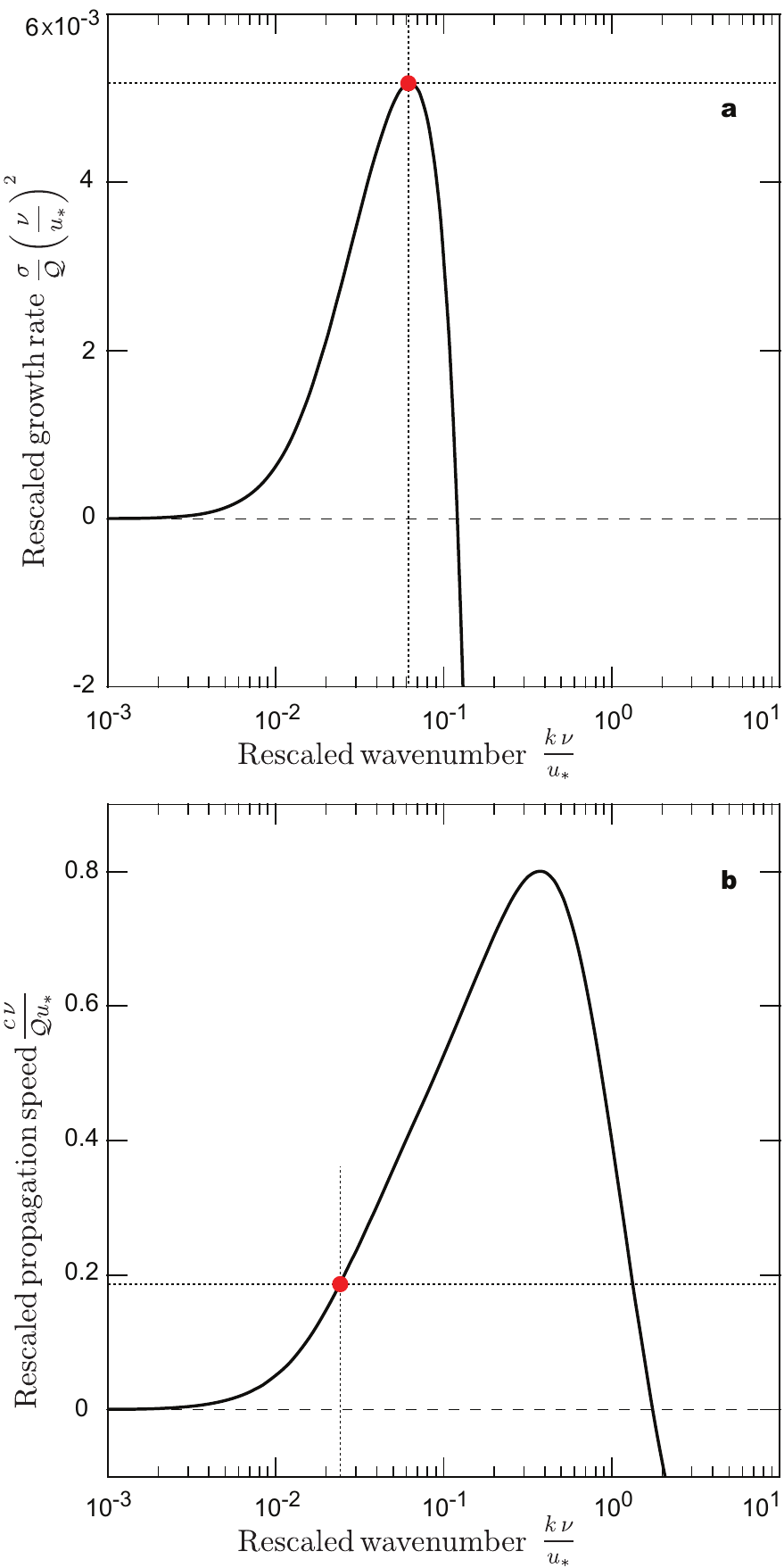}}
\noindent
{{\bf Figure S7.}
Dispersion relation: dimensionless growth rate (\textbf{a}) and propagation speed (\textbf{b}) as functions of the rescaled wavenumber $k\nu/u_*$, computed at perihelion for $d=4$~mm, with a saturation length  $L_{\rm sat}/d=24$. This corresponds to the neck (Hapi) region, where the observed emergent ripple wavelength $\lambda$ is around $7$~m. The corresponding most unstable mode (red dot) is at $k\nu/u_* \simeq 0.06$. Vapor viscosity and shear velocity are respectively $\nu \simeq 5$~m$^2$/s and $u_* \simeq 70$~m/s, respectively. With a reference sediment flux $\mathcal{Q} \simeq 4 \, 10^{-6}$~m$^2$/s, the growth rate of this mode is $\sigma_m \simeq 5.2 \, 10^{-3}  \mathcal{Q} (u_*/\nu)^2 \simeq 5 \, 10^{-6}$~s$^{-1}$. Mature ripples at a wavelength of $18$~m ($k\nu/u_* \simeq 0.024$) propagate at a velocity $c \simeq 0.18 \, \mathcal{Q} u_*/\nu \simeq 10^{-5}$~m/s, i.e. over $\simeq 10$~m for the time during sediment transport occurs $\simeq 10^6$~s.
}

\newpage
{\tiny
\begin{center}
\begin{tabular}{|l|c|c|c|c|c|c|}
\hline
Photo name 		& $N$	& $\lambda$ 	&  $t$   		& Region	& Fig. 	\\
Web link for picture 	&		& (m)		& ($10^7$ s)  	& 		& 		\\
\hline
\hline
Comet\_from\_9\_m								& 		& 		& -2.36	& Ma'at	& 	2B\\
\href{http://www.esa.int/spaceinimages/Images/2015/07/Comet_from_9_m}{\texttt{www.esa.int/spaceinimages/Images/2015/07/Comet\_from\_9\_m}} & & & & & \\
\hline
\hline
Comet\_from\_67.4\_m								& 1		& 27		& -2.36 	& Ma'at	& 	\\
\href{http://www.esa.int/spaceinimages/Images/2015/07/Comet_from_67.4_m}{\texttt{www.esa.int/spaceinimages/Images/2015/07/Comet\_from\_67.4\_m}} & & & & & \\
\hline
Comet\_from\_67.4\_m								& 5		& 4		& -2.36	& Ma'at	& 	\\
\href{http://www.esa.int/spaceinimages/Images/2015/07/Comet_from_67.4_m}{\texttt{www.esa.int/spaceinimages/Images/2015/07/Comet\_from\_67.4\_m}} & & & & & \\
\hline
NAC\_2016-04-13T15.17.54.813Z\_ID10\_1397549800\_F22	& 11		& 16.5	& 2.09	& Ma'at	&	\\
\href{https://planetgate.mps.mpg.de/Image_of_the_Day/public/OSIRIS_IofD_2016-04-19.html}{\texttt{planetgate.mps.mpg.de/Image\_of\_the\_Day/public/OSIRIS\_IofD\_2016-04-19.html}} & & & & & \\
\hline
NAC\_2016-01-10T15.58.51.484Z\_ID10\_1397549008\_F22	& 15		& 17.5	& 1.29	& Ma'at	&	\\
\href{https://planetgate.mps.mpg.de/image_of_the_day/public/OSIRIS_IofD_2016-01-18.html}{\texttt{planetgate.mps.mpg.de/image\_of\_the\_day/public/OSIRIS\_IofD\_2016-01-18.html}} & & & & & \\
\hline
NAC\_2016-03-05T11.36.49.540Z\_ID30\_1397549100\_F24	& 1		& 20		& 1.76	& Ma'at	&	S1B\\
\href{https://planetgate.mps.mpg.de/image_of_the_day/public/OSIRIS_IofD_2016-03-12.html}{\texttt{planetgate.mps.mpg.de/image\_of\_the\_day/public/OSIRIS\_IofD\_2016-03-12.html}} & & & & & \\
\hline
NAC\_2016-05-21T11.41.59.934Z\_ID20\_1397549001\_F22	& 1		& 20		& 1.56	& Ma'at	&	\\
\href{https://planetgate.mps.mpg.de/image_of_the_day/public/osiris_iofd_2016-05-23.html}{\texttt{planetgate.mps.mpg.de/image\_of\_the\_day/public/osiris\_iofd\_2016-05-23.html}} & & & & & \\
\hline
NAC\_2016-01-17T06.55.38.746Z\_ID10\_1397549500\_F22	& 11		& 25		& 1.35	& Ma'at	&	1B\\
\href{https://planetgate.mps.mpg.de/Image_of_the_Day/public/OSIRIS_IofD_2016-01-22.html}{\texttt{planetgate.mps.mpg.de/Image\_of\_the\_Day/public/OSIRIS\_IofD\_2016-01-22.html}} & & & & & S1A\\
\hline
NAC\_2016-05-21T11.41.59.934Z\_ID20\_1397549001\_F22	& 		& 		& 2.4 	& Ma'at	&	1A\\
\href{https://planetgate.mps.mpg.de/Image_of_the_Day/public/OSIRIS_IofD_2016-05-23.html}{\texttt{planetgate.mps.mpg.de/Image\_of\_the\_Day/public/OSIRIS\_IofD\_2016-05-23.html}} & & & & & S1B\\
\hline
NAC\_2016-01-17T06.55.38.746Z\_ID10\_1397549500\_F22	& 4		& 20		& 1.35	& Hapi	&	\\
\href{https://planetgate.mps.mpg.de/Image_of_the_Day/public/OSIRIS_IofD_2016-01-22.html}{\texttt{planetgate.mps.mpg.de/Image\_of\_the\_Day/public/OSIRIS\_IofD\_2016-01-22.html}} & & & & & \\
\hline
NAC\_2016-02-27T15.33.24.581Z\_ID30\_1397549500\_F22	& 2		& 16		& 1.70	& Hapi	&	\\
\href{https://planetgate.mps.mpg.de/Image_of_the_Day/public/OSIRIS_IofD_2016-03-05.html}{\texttt{planetgate.mps.mpg.de/Image\_of\_the\_Day/public/OSIRIS\_IofD\_2016-03-05.html}} & & & & & \\
\hline
NAC\_2016-06-15T21.49.20.545Z\_ID10\_1397549600\_F22	& 3		& 17.5	& 2.64	& Hapi	&	\\
\href{https://planetgate.mps.mpg.de/image_of_the_day/public/OSIRIS_IofD_2016-06-24.html}{\texttt{planetgate.mps.mpg.de/image\_of\_the\_day/public/OSIRIS\_IofD\_2016-06-24.html}} & & & & & \\
\hline
NAC\_2016-06-15T21.49.20.545Z\_ID10\_1397549600\_F22	& 6		& 12		& 2.64	& Hapi	&	\\
\href{https://planetgate.mps.mpg.de/image_of_the_day/public/OSIRIS_IofD_2016-06-24.html}{\texttt{planetgate.mps.mpg.de/image\_of\_the\_day/public/OSIRIS\_IofD\_2016-06-24.html}} & & & & & \\
\hline
NAC\_2016-06-15T21.49.20.545Z\_ID10\_1397549600\_F22	& 8		& 7		& 2.64	& Hapi	&	\\
\href{https://planetgate.mps.mpg.de/image_of_the_day/public/OSIRIS_IofD_2016-06-24.html}{\texttt{planetgate.mps.mpg.de/image\_of\_the\_day/public/OSIRIS\_IofD\_2016-06-24.html}} & & & & & \\
\hline
ROS\_CAM1\_20141024T180435\_P					& 12		& 7		& -2.52	& Hapi	&	1B\\
\href{http://imagearchives.esac.esa.int/picture.php?/8905/category/64}{\texttt{imagearchives.esac.esa.int/picture.php?/8905/category/64}} & & & & & \\
\hline
ROS\_CAM1\_20141024T180435\_P					& 3		& 16		& -2.52	& Hapi	&	1B\\
\href{http://imagearchives.esac.esa.int/picture.php?/8905/category/64}{\texttt{imagearchives.esac.esa.int/picture.php?/8905/category/64}} & & & & & \\
\hline
NAC\_2016-02-27T06.58.40.552Z\_ID10\_1397549600\_F22	& 15		& 7.5		& 1.70	& Ash	&	\\
\href{https://planetgate.mps.mpg.de/Image_of_the_Day/public/OSIRIS_IofD_2016-03-01.html}{\texttt{planetgate.mps.mpg.de/Image\_of\_the\_Day/public/OSIRIS\_IofD\_2016-03-01.html}} & & &  & & \\
\hline
NAC\_2016-02-27T06.58.40.552Z\_ID10\_1397549600\_F22	& 6		& 12.5	& 1.70	& Ash	&	\\
\href{https://planetgate.mps.mpg.de/Image_of_the_Day/public/OSIRIS_IofD_2016-03-01.html}{\texttt{planetgate.mps.mpg.de/Image\_of\_the\_Day/public/OSIRIS\_IofD\_2016-03-01.html}} & &  & & & \\
\hline
NAC\_2016-06-06T18.19.07.691Z\_ID20\_1397549100\_F22	& 7		& 12		& 2.57	& Ash	&	\\
\href{https://planetgate.mps.mpg.de/image_of_the_day/public/OSIRIS_IofD_2016-06-08.html}{\texttt{planetgate.mps.mpg.de/image\_of\_the\_day/public/OSIRIS\_IofD\_2016-06-08.html}} & &  & & & \\
\hline
NAC\_2016-06-06T18.19.07.691Z\_ID20\_1397549100\_F22	& 10		& 9		& 2.57	& Ash	&	\\
\href{https://planetgate.mps.mpg.de/image_of_the_day/public/OSIRIS_IofD_2016-06-08.html}{\texttt{planetgate.mps.mpg.de/image\_of\_the\_day/public/OSIRIS\_IofD\_2016-06-08.html}} & &  & & & \\
\hline
Rosetta's last image									& 		& 		& 3.58	& Ma'at	& 2D	\\
\href{http://www.esa.int/spaceinimages/Images/2016/09/Rosetta_s_last_image}{\texttt{www.esa.int/spaceinimages/Images/2016/09/Rosetta\_s\_last\_image}} & &  & & & \\
\hline
\end{tabular}
\end{center}
}
{{\bf Table S1.}
Ripple crest-to-crest distance measured on pictures of different regions of 67P. $N+1$ is the number of successive ripple crests identified on the picture. $\lambda$ is the average value of their distance. $t$ is the time to perihelion (13 Aug. 2015), counted positive (negative) after (before) it. The last column gives the figure number where the corresponding picture has been used.
}

\end{document}